\documentclass[%
 reprint,
 amsmath,amssymb,
 aps,
pra,
superscriptaddress,
]{revtex4-2}

\usepackage{graphicx}
\usepackage{dcolumn}
\usepackage{bm}
\usepackage{verbatim}
\usepackage{color}

\begin{document}

\title{Preparing squeezed spin  states in a spin-mechanical hybrid system with silicon-vacancy centers }%

\author{Bo Li}
\affiliation{Department of Physics and Astronomy, Purdue University, West Lafayette, IN 47907, USA}
\affiliation{Shaanxi Province Key Laboratory of Quantum Information and Quantum Optoelectronic Devices,
             Department of Applied Physics, Xi'an Jiaotong University, Xi'an 710049, China}%
\author{Xiaoxiao Li}
\author{Pengbo Li}
\email{lipengbo@mail.xjtu.edu.cn}
\affiliation{Shaanxi Province Key Laboratory of Quantum Information and Quantum Optoelectronic Devices,
             Department of Applied Physics, Xi'an Jiaotong University, Xi'an 710049, China}%
\author{Tongcang Li}
\email{tcli@purdue.edu}
\affiliation{Department of Physics and Astronomy, Purdue University, West Lafayette, IN 47907, USA}
\affiliation{School of Electrical and Computer Engineering, Purdue University, West Lafayette, Indiana 47907, USA}
\affiliation{Purdue Quantum Science and Engineering Institute, Purdue University, West Lafayette, Indiana 47907, USA}
\affiliation{Birck Nanotechnology Center, Purdue University, West Lafayette, Indiana 47907, USA}

\date{\today}

\begin{abstract}
We present and analyze an effective scheme for preparing squeezed spin states
in a novel spin-mechanical hybrid device, which is realized by a single crystal diamond  waveguide
with built-in silicon-vacancy (SiV) centers.  
After studying the strain couplings between the SiV spins and the propagating phonon modes,
we show that long-range spin-spin interactions  can be achieved under large detuning condition. 
We model these nonlinear spin-spin couplings with an effective one-axis twisting Hamiltonian,
and find that  the system can be steered to the squeezed spin states in the practical situations.
This work may have interesting applications in high-precision metrology and quantum information.
\end{abstract}

\maketitle

\section{introduction}

Solid defects in diamond including nitrogen-vacancy (NV) \cite{PR-528-1}, silicon-vacancy (SiV) \cite{PhysRevLett.112.036405,PhysRevLett.113.113602,Sipahigil847}, 
and germanium vacancy (GeV) \cite{PhysRevLett.118.223603} color centers have attracted great interest in recent years.
With the advantage of long spin-coherence times, they provide promising platforms for quantum sensing
and quantum information processing.  A most prominent example is the NV center,
for which the detection of weak signals \cite{Science-364-973}, design of quantum logic gates
\cite{Natrue-514-72,PhysRevA.99.022319,xu2019quantum}, and storage or transfer of quantum information \cite{PhysRevA.84.010301,PhysRevA.96.032342,PhysRevA.91.042307,PhysRevApplied.11.044026,yin2015hybrid}
have been investigated theoretically and experimentally.
However, SiV centers are increasingly recognized as more attractive in some cases because they
provide favorable optical properties such as bright, spectrally stable emission and very large strain susceptibility.
In particular, recent demonstrations of long  spin-coherence times ($\sim$10 ms at low temperature) \cite{PhysRevLett.119.223602},
and spin-photon interfaces \cite{2016NatCo_Becker}, as well as a variety of quantum controls \cite{PhysRevLett.113.263602,PhysRevLett.120.053603,2017NatCo_Pingault,Jahnke_2015,Nature.Comm-SiV-strain,
PhysRevB.94.214115,PhysRevB.97.205444,arxiv-1901-04650}, make the SiV center a good memory qubit.
 More importantly, the very recent research shows that the strong strain couplings
 between the individual SiV center spins and the propagating phonon modes
can be achieved \cite{PhysRevLett.120.213603}.
 This may set a base for manipulating the quantum states of SiV centers
 via the spin-phonon couplings.

As spin-based quantum technology evolves, there has been a considerable effort to manipulate
the solid defect centers with mechanical driving \cite{Lee_2017}.
Generally speaking, the static strain will generate energy shifts in the ground states,
and the resonant a.c. strain field can coherently drive electron spin transitions.
Thus far, the strong mechanical driving spin transitions with NV centers or divacancy centers
in silicon carbide have been experimentally demonstrated  \cite{NatPhy_NV_AC,PhysRevLett.113.020503,2018arXiv180410996W}.
However, as the sublevels of the ground state in NV centers are defined within the same orbital,
the strain susceptibility is very weak ($\sim 10^{9}$ Hz/strain) \cite{2018arXiv180410996W,NL-17-NV,2018arXiv180410996W,2013NatC-4E1819F}.
By contrast, for SiV centers, because the ground states are defined in distinct orbital branches,
they provide much larger strain susceptibility ($\sim 10^{15}$ Hz/strain)  \cite{Nature.Comm-SiV-strain,Jahnke_2015,PhysRevB.94.214115,PhysRevB.97.205444},
and are more attractive to structure  spin-mechanical hybrid devices. 
To further explore the potential of the spin-mechanical hybrid quantum systems,
it is very appealing to steer the collective spins to
the entangled states or the spin squeezed states (SSSs).

Arising from quantum correlation of collective spin systems, spin squeezing means
partly cancel out fluctuations in one direction at the expense of those enhanced
in the other directions \cite{PhysRevA.47.5138,MA201189,Jin_2009}.
Providing the ability to interrogate a physical system and precisely
measure its observables, SSSs are fundamental to quantum information and high-precision metrology
\cite{RevModPhys.89.035002,Science-355-620,arXiv-1801-00042}.
After decades of development, several schemes for preparing the SSSs with large atomic ensembles have been demonstrated.
For instance, one can directly transfer the SSSs from
the optical field to the atomic ensembles \cite{RevModPhys.82.1041,2001Natur.413.400J,PhysRevA.62.063812}.
However, in this protocol, the degree of spin squeezing is limited by the quality of the
input optical field.
Other promising schemes include quantum non-demolition measurement \cite{PhysRevLett.99.163002,PhysRevLett.110.163602},
or cavity squeezing \cite{PhysRevLett.101.073601,Shindo_2003,PhysRevA.73.053817,PhysRevLett.110.120402}.
As for spin-mechanical hybrid architectures, the spin-phonon couplings can be utilized for
driving the collective spins.
Specifically, an effective scheme for preparing the SSSs with NV centers 
through mechanical driving 
has been theoretically studied in Ref. \cite{PhysRevLett.110.156402}.
However, because the coupling interactions between NV spins and mechanical phonon modes are very weak,
further experimental demonstrations are needed to verify its feasibility.

In this work, we propose an experimentally feasible scheme
for generating the SSSs with SiV centers in a novel  spin-mechanical hybrid device.
In particular, the setup under consideration is realized by a diamond
waveguide with SiV spins embedded.
We first describe the strain couplings between the SiV centers and the propagating phonons,
and show that the long-range  spin-spin interactions  can be obtained under  large detuning condition.
We model these  nonlinear  spin-spin couplings  with an effective one-axis twisting  Hamiltonian,
and find that the system can be steered to SSSs as time envolves.
We further study the effects of decoherence processes with numerical simulations,
showing that our scheme is feasible in the realistic conditions.
The benefits of this spin-mechanical hybrid device  are quite diverse.
For one thing, SiV centers provide long spin-coherence times,
and their ground states can be tuned by external magnetic fields via Zeeman effect.
For another, the recent research shows that the spin-phonon couplings can reach $2\pi \times 14$ MHz
even for single SiV centers \cite{PhysRevLett.120.213603}, which far exceeds the decoherence rates.
Because the collective spin-phonon couplings are used for spin squeezing in our scheme,
the coupling strength would further be enhanced.
Apart from SSSs, there are more potential applications utilizing this hybrid quantum device,
such as the entanglement detections \cite{Natur-409-63}.

\section{The Setup}

\subsection{Electronic structure of the SiV centers}

\begin{figure}[b]
\includegraphics[width=8.7cm]{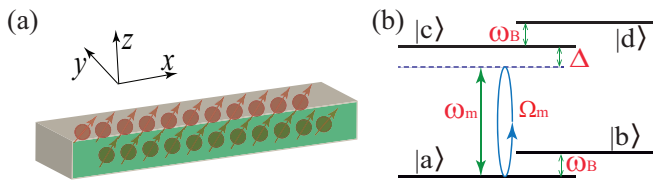}
\caption{\label{fig_1} Schematic design and operation of the setup.
(a) The SiV center ensemble is embedded in a quasi-1D diamond waveguide.
(b) In the presence of static  magnetic field,
the  ground states of each SiV center can be mapped to a four-level system,
and the spin transitions induced by the mechanical phonon modes
 are between $|a\rangle \leftrightarrow |c\rangle $ and
$|b\rangle \leftrightarrow |d\rangle $.
In the case without magnetic field,
the SiV ground states   simplify to a two level system, denoted as $|a\rangle$
and $|c\rangle$.}
\end{figure}

As shown schematically in Fig.~1(a),
the setup under consideration is realized by a single crystal diamond 
waveguide with built-in SiV centers.
In particular, we consider a  waveguide with size $(l,w,t)$,
and focus on the quasi-one-dimensional (1D) condition $l\gg \{w,t\}$.
We suppose the symmetry axis of the SiV centers is
along the $z$ axis, and the compression phonon modes supported by the waveguide is along the $x$ axis.
In the diamond lattice, the SiV center is a point defect in which a silicon atom is positioned
between two adjacent missing carbon atoms. The electronic ground state of SiV center consists of an unpaired
hole with spin $S=1/2$, which can occupy one of the two degenerate orbital states $|e_{x}\rangle $
 or $|e_{y}\rangle$. In the presence of an external magnetic field $\vec{B}=B_{z}\vec{e}_{z}$,
the SiV center spin is captured by Hamiltonian (let $\hslash=1$)
\begin{equation}
\hat{H}_{SiV}=-\lambda _{so}\hat{L}_{z}\hat{S}_{z}+\hat{H}_{JT}+f\gamma
_{L}B_{z}\hat{L}_{z}+\gamma _{s}\vec{B}\cdot \hat{\vec{S}},
\end{equation}
where $\lambda _{SO}>0$ is the spin-orbit coupling constant,
$\hat{L}_{z}$ and $\hat{S}_{z}$   are the projections of the dimensionless angular
momentum and spin operators $\hat{\vec{L}}$ and $\hat{\vec{S}}$ onto the symmetry $(z)$
axis of the center, $\hat{H}_{JT}$ is the Jahn-Teller (JT) effect,
$f\approx 0.1$ is the reduced orbital Zeeman effect, and
$\gamma _{L}(\gamma _{s})$ is the orbital (spin) gyromagnetic ratio.

In weak magnetic field environment, the SiV spin Hamiltonian (1) is dominated by
the spin-orbit coupling, with a constant $\lambda _{SO}\simeq 2\pi \times 45$ GHz.
This coupling interaction splits the ground state manifold into two lower
states $\{|e_{-},\downarrow \rangle ,|e_{+},\uparrow \rangle \}$ and two upper states
$\{|e_{+},\downarrow \rangle ,|e_{-},\uparrow \rangle \}$,
 where $|e_{\pm }\rangle =(|e_{x}\rangle \pm i|e_{y}\rangle )/\sqrt{2}$  are eigenstates of the angular
momentum operator, $\hat{L}_{z}|e_{\pm }\rangle =\pm |e_{\pm }\rangle $. 
The reduced orbital Zeeman term can be neglected in our discussion,
as it does not affect the results.
In addition, the JT interaction $\hat{H}_{JT}$ with strength
$\Upsilon =\sqrt{\Upsilon _{x}^{2}+\Upsilon _{y}^{2}}\ll \lambda _{SO}$
results in a mixing between the orbital angular momentum states $|e_{+}\rangle $ and $|e_{-}\rangle $.
Putting everything together, diagonalization of the SiV spin Hamiltonian (1) leads to a four-level system
\cite{PhysRevLett.120.213603,arxiv-1901-04650,PhysRevB.94.214115,PhysRevB.97.205444},
 with eigenstates
\begin{eqnarray}
|a\rangle  &=&(\cos\theta |e_{x}\rangle -i\sin\theta e^{-i\phi }|e_{y}\rangle
)|\downarrow \rangle ,  \nonumber \\
|b\rangle  &=&(\cos\theta |e_{x}\rangle +i\sin\theta e^{i\phi }|e_{y}\rangle
)|\uparrow \rangle ,  \nonumber \\
|c\rangle  &=&(\sin\theta |e_{x}\rangle +i\cos\theta e^{-i\phi }|e_{y}\rangle
)|\downarrow \rangle ,  \nonumber \\
|d\rangle  &=&(\sin\theta |e_{x}\rangle -i\cos\theta e^{i\phi }|e_{y}\rangle
)|\uparrow \rangle ,
\end{eqnarray}
where
$\tan(\theta )=(2\Upsilon _{x}+\Delta )/\sqrt{\lambda
_{SO}^{2}+4\Upsilon _{y}^{2}}$, and $\tan(\phi )=2\Upsilon _{y}/\lambda _{SO}$.
The corresponding eigenenergies read $\omega _{a,b}=-(D \pm \omega _{B})/2$ and
$\omega _{c,d}=(D \mp \omega _{B})/2$, where
 $D=\sqrt{\lambda _{SO}^{2}+4\Upsilon ^{2}}\approx 2\pi \times 46$ GHz
is the intrinsic splitting of the ground states, and
 $\omega _{B}=\gamma _{s}B_{z}$ is the Zeeman energy. 　
Remarkably, because $\Upsilon _{x,y}\ll \lambda _{SO}$, we can ignore the small
distortions of the orbital states induced by the JT effect, then obtain a simplified expression
$|a\rangle \approx |e_{-},\downarrow \rangle ,|b\rangle \approx
|e_{+},\uparrow \rangle ,|c\rangle \approx |e_{+},\downarrow \rangle $
and $|d\rangle \approx |e_{-},\uparrow \rangle $.

\subsection{Spin-phonon couplings}

In this spin-mechanical hybrid device, the time-dependent displacement field
associated with internal compression modes of the waveguide
affect the Coulomb energy of the electronic states,
 which induces the spin-phonon couplings.
 In the quasi-1D condition,  we assume that the
orbital degrees of freedom of the SiV centers
are coupled  to a continuum of   compression  
modes propagating along the $x$ axis \cite{PhysRevLett.120.213603,arxiv-1901-04650}.
After quantization, the  displacement field associated with the  propagating phonon modes
can be expressed as
\begin{equation}
\vec{u}(\vec{r})=\sum\limits_{k,n}Q_{n,k}^{0}\vec{u}_{n,k}^{\perp }(y,z)(%
\hat{b}_{n,k}e^{ikx}+\hat{b}_{n,k}^{\dagger }e^{-ikx}),
\end{equation}
where $Q_{n,k}^{0}$ is the effective zero point fluctuations, $\vec{u}_{n,k}^{\perp }(\vec{r},t)$
is the transverse mode profile,  $\hat{b}_{n,k}$ and $\hat{b}_{n,k}^{\dagger }$  are the annihilation
and creation operators of the propagating phonons with eigen frequency $\omega _{n,k}$,
$k$ is the wave vector, and $n$ is the band index (see Appendix A and B). As a result,
the  spin-mechanical dynamics  for the whole system can be described by Hamiltonian
\begin{eqnarray}
\hat{H}_{sys} &=&\sum\limits_{j}\hat{H}_{SiV}^{(j)}+\sum\limits_{n,k}\omega
_{n,k}\hat{b}_{n,k}^{\dagger }\hat{b}_{n,k}  \nonumber \\
&&+\sum\limits_{j,n,k}(g_{n,k}^{j}\hat{\sigma}_{-}^{j}\hat{b}_{n,k}^{\dagger
}e^{-ikx_{j}}+H.c.),
\end{eqnarray}
where $j$ labels the SiV centers located at positions $\vec{r}_{j}(x_{j},y_{j},z_{j})$,
 $g_{n,k}^{j}$ is the spin-phonon coupling constant,
and $\hat{\sigma}_{-}^{j}=(\hat{\sigma}_{+}^{j})^{\dagger }=|a^{j}\rangle
\langle c^{j}|+|b^{j}\rangle \langle d^{j}|$ is the
spin-conserving lowering operator.

For a waveguide with small size, because the phonon modes are well separated in
frequency ($\Delta \omega _{n}\geq 2\pi \times 50$ MHz),
the defect centers are coupled predominantly to only one compression mode of the waveguide (see Appendix C).
As a result, the mechanical phonon modes
can be simplified to a single standing-wave mode $\hat{b}$ with frequency
$\omega _{m}\sim D\approx 2\pi \times 46$ GHz.
In particular, depending on the local strain tensor $\varepsilon _{ij}$
and the mode profile $\vec{u}_{n,k}^{\perp }(\vec{r},t)$,
the spin-phonon coupling constant for single SiV center can be estimated as \cite{PhysRevLett.120.213603,arxiv-1901-04650}
\begin{equation}
g\simeq \frac{d}{v_{l}}\sqrt{\frac{\hbar \omega _{m}}{2\rho lwt}},
\end{equation}
where $d/2\pi \sim 10^{15}$ Hz/strain
is the strain sensitivity,
$v_{l}$ is the speed of the longitudinal compression modes in the waveguide, and $\rho $ is the  density of diamond.

For the purpose of simplification, we consider the case $B_{z}\rightarrow 0$.
Then the SiV center spins can be mapped to a two-level system (i.e., spin-$\frac{1}{2}$ particles)
denoted as $|a^{j}\rangle $ and $|c^{j}\rangle $,
and the spin-conserving lowering operator
simplifies to $\hat{\sigma}_{-}^{j}\rightarrow |a^{j}\rangle \langle c^{j}|$.
 We further introduce the collective spin operators
 $\hat{J}_{-}=(\hat{J}_{+})^{\dagger }=\sum\nolimits_{j=1}^{N}|a^{j}\rangle
\langle c^{j}|$, $\hat{J}_{x}=\frac{1}{2}(\hat{J}_{-}+\hat{J}_{+})$, and $\hat{J}_{y}=\frac{%
i}{2}(\hat{J}_{-}-\hat{J}_{+})$, which satisfy the usual angular momentum commutation relations
\begin{equation}
\lbrack \hat{J}_{\alpha },\hat{J}_{\beta }]=i\varepsilon _{\alpha \beta
\gamma }\hat{J}_{\gamma },\text{ \ \ }[\hat{J}_{+},\hat{J}_{-}]=2\hat{J}_{z},%
\text{ \ \ }[\hat{J}_{z},\hat{J}_{\pm }]=\pm \hat{J}_{\pm },
\end{equation}
where  $\varepsilon _{\alpha \beta \gamma }$ is the Levi-Civita symbol,
and $\hat{J}_{z}=\frac{1}{2}\sum\nolimits_{j=1}^{N}|c^{j}\rangle \langle
c^{j}|-|a^{j}\rangle \langle a^{j}|$.
The Hamiltonian of the
whole system then can be represented as
\begin{equation}
\hat{H}_{sys}=\omega _{m}\hat{b}^{\dagger }\hat{b}+\omega _{s}\hat{J}_{z}+g_{e}(%
\hat{b}^{\dagger }\hat{J}_{-}+\hat{b}\hat{J}_{+}),
\end{equation}
where $g_{e}\simeq \sqrt{N}g$ is the collective coupling strength, and
$\omega _{s}\sim D\approx 2\pi \times 46$ GHz
is the resonance frequency of the SiV spins.
Hamiltonian (7) describes a Tavis-Cummings type interaction
between an ensemble of SiV spins and a single mechanical phonon mode.

\subsection{Effective model}

We now consider the spin-phonon couplings in the large detuning condition $g_{e}\ll \Delta =|\omega _{s}-\omega _{m}|$.
In a frame rotating at the mechanical frequency $\omega _{m}$, the system can be described by Hamiltonian
\begin{equation}
\hat{H}_{I}=\Delta \hat{J}_{z}+g_{e}(\hat{b}^{\dagger }\hat{J}_{-}+\hat{b}%
^{\dagger }\hat{J}_{+}),
\end{equation}
with $\Delta$ the detuning of the spin-phonon couplings.
To study the phonon-mediated spin-spin coupling interactions,
we  introduce the Fr\"{o}lich-Nakajima transform
\begin{equation}
\hat{H}_{eff}=e^{-R}\hat{H}e^{R}, 
\end{equation}
where $R=g_{e}(\hat{b}^{\dagger }\hat{J}_{-}-\hat{b}J_{+})/\Delta $. 
By performing the  Fr\"{o}lich-Nakajima transform to $\hat{H}_{I}$,
and using $\hat{J}_{+}\hat{J}_{-}=\hat{J}^{2}-\hat{J}_{z}^{2}+\hat{J}_{z}$, 
to order $(g_{e}/\Delta )^{2}$ we obtain an effective Hamiltonian \cite{MA201189,PhysRevLett.110.156402}
\begin{equation}
\hat{H}_{eff}\simeq (\Delta +\lambda +2\lambda \hat{b}^{\dagger }\hat{b})%
\hat{J}_{z}-\lambda \hat{J}_{z}^{2},
\end{equation}
where $\lambda =g_{e}^{2}/\Delta $ is the coupling strength of the effective spin-spin interactions.
In our system, these nonlinear spin-spin couplings are
induced by virtual excitations of the mechanical modes.
We note that when ignoring the  fluctuations of the phonon number $n=\hat{b}^{\dagger }\hat{b}$,
the effective Hamiltonian (10) simplifies to a standard one-axis twisting model.

\section{Experimental Feasibility}

We now consider the experimental feasibility of our scheme and the appropriate parameters to achieve strong
couplings. In view of the recent progress in the
fabrication of high-quality diamond structures \cite{domaind-Burek,2014NC-5-3638T,2012ApPhL.101p3505O},
and demonstrations of strain-induced control
of defects \cite{PhysRevLett.120.213603,arxiv-1901-04650,PhysRevB.94.214115,
PhysRevB.97.205444,Lee_2017,NatPhy_NV_AC,PhysRevLett.113.020503,2018arXiv180410996W},
 the proposed scheme could realistically be
implemented with current experimental techniques.
For the diamond waveguides, the material properties are $E=1050$ Gpa, $\nu =0.2$, and $\rho =3500$ kg/m$^{3}$.
 The group velocities for a longitudinal compression and a
transverse flexural mode are $v_{l}=1.7\times 10^{4}$ m/s
and  $v_{t}=0.73\times 10^{4}$ m/s, respectively. We consider
a quasi-1D diamond waveguide with  size  $(0.1,0.1,20)$ $\mu $m$^{3}$,
then the coupling constant can be estimated as $g\sim 3.4$ MHz for individual SiV centers.
For a SiV ensemble with spin number $N \sim 1000$, a collective spin-phonon coupling
$g_{e}\simeq \sqrt{N}g\sim 100$ MHz can be reached.
In  the large detuning condition $\Delta \sim 10g_{e}$,
the effective spin-spin coupling strength can be estimated as $\lambda \simeq g_{e}^{2}/\Delta \sim 10$ MHz.
By adjusting the size and spin number of the diamond waveguide,
these  spin-spin couplings could further be enhanced. 

In the practical situations, a quality factor $Q\sim 5\times 10^{4}$
for the mechanical phonon modes is realistic, which  results in a mechanical decay rate
$\gamma _{m}\simeq \omega _{m}/Q\sim 2\pi \times 1$ MHz.
In the low-temperature environment $T\sim 100$ mK,
the spin-coherence time of SiV centers  can reach $\sim 10$ ms with dynamical decoupling \cite{PhysRevLett.119.223602}.
Even taking into account the surface effect on SiV centers' decoherence,
the spin dephasing rate is still expected to be
about $\gamma _{s}\sim 2\pi \times 100$ kHz.
The thermal phonon number $n_{th}\simeq(e^{\hslash  \omega _{m}/k_{B}T}-1)^{-1}$
 is far below $1$ in such low  temperature.
 With these realistic coupling (decay) parameters,
we perform numerical simulation for the dynamics of the system,  
the result is displayed in Fig 3(a).
We find that in the practical situations, the effective spin-spin couplings can
 dominate the decoherence processes, and steer the SiV centers to SSSs as time evolves.

\section{Spin Squeezing }
\subsection{The ideal case}

We first discuss the spin squeezing of SiV centers in this spin-mechanical hybrid device
in the ideal case,
then turn to the decoherence effects in the realistic condition.
Specifically, the spin squeezing in our system can be introduced by the Heisenberg uncertainty relation.
As the  collective spin operators $\{\hat{J}_{x},\hat{J}_{y},\hat{J}_{z}\}$ 
obey SU(2) algebra, the associated uncertainty
relation $(\Delta \hat{J}_{\alpha })^{2}(\Delta \hat{J}_{\beta })^{2}\geq |\langle
\hat{J}_{\gamma }\rangle |^{2}/4$ can be satisfied.
Using optical pumping and microwave spin manipulation,
we can firstly preparing the SiV centers to the
coherent spin state $|\psi _{0}\rangle $ along the $x$ axis,
which satisfies  $\hat{J}_{x}|\psi _{0}\rangle =J|\psi _{0}\rangle $, where $J=N/2$.
As the coherent spin state is also known as minimum-uncertainty state,
 we have  $(\Delta \hat{J}_{y})^{2}=(\Delta \hat{J}_{z})^{2}=|\hat{J}_{x}|/2=J/2$,
 with  $J/2$ the standard quantum limit.

The spin squeezing is defined if the variance of a spin component normal to the mean spin is
smaller than the standard quantum limit, i.e., $(\Delta \hat{J}_{\vec{n}_{\perp }})^{2}<J/2$.
In order to describe the degree of spin squeezing, we
introduce the definition of spin squeezing parameter given by
 Kitagawa and Ueda \cite{PhysRevA.47.5138}
\begin{equation}
\xi ^{2}=\frac{4(\Delta \hat{J}_{\vec{n}_{\perp }}^{2})_{\min}}{N}
\end{equation}
where
$\vec{n}_{\perp }=\sin\alpha\cdot \vec{y}+\cos\alpha\cdot \vec{z}$
refers to an axis perpendicular to the mean-spin direction $\vec{x}$,
and the minimization is over all directions $\vec{n}_{\bot }$.
Note that the spin coherent states and the SSSs can be 
distinguished by  $\xi ^{2}=1$ or $\xi ^{2}<1$.
Further more, as theoretically studied
in Ref. \cite{PhysRevA.47.5138,MA201189,Jin_2009}, for spin-$\frac{1}{2}$ particles the minimization spin
uncertainty can also be represented as
\begin{equation}
(\Delta \hat{J}_{\vec{n}_{\perp }}^{2})_{\min}=\frac{1}{2}(A-\sqrt{%
B^{2}+4C^{2}})
\end{equation}
where $A=\langle \hat{J}%
_{y}^{2}+\hat{J}_{z}^{2}\rangle $, $B=\langle \hat{J}_{y}^{2}-\hat{J}%
_{z}^{2}\rangle $, and $C=\frac{1}{2}\langle \hat{J}_{y}\hat{J}_{z}+\hat{J}_{z}\hat{J}_{y}\rangle $. %

We now consider  preparing the SSSs of SiV centers in our system using the
dynamic evolution of the system.
 Under large detuning condition, we show that the effective spin-spin interactions can be achieved
through the spin-phonon couplings.
We model these long-range spin-spin interactions with the effective Hamiltonian (10),
and find that a  standard one-axis twisting model can be obtained when ignoring the phonon fluctuations.
We remark that the one-axis twisting interaction works in analogy to the squeezing operator,
which can steer the system to the SSSs with properly initial states. 
In particular, this spin squeezing protocol
 has been implemented in BEC via atomic collisions \cite{2008Natur-409-63S,2010Natur.464.1165G,2010Natur.464.1170R}
and in atomic ensembles via large detuned atom-field
interactions \cite{PhysRevLett.99.163002}.
Moreover, the analytic solutions of the standard one-axis twisting model has been studied in
Ref. \cite{PhysRevA.47.5138,MA201189,Jin_2009}. In our system,
as time evolves, the minimization spin uncertainty
in the short-time limit ($\lambda t\ll 1$) and large particle number ($J\gg 1$)
can be calculated as
\begin{equation}
\ (\Delta \hat{J}_{\vec{n}_{\perp }}^{2})_{\min}\simeq \frac{J}{2}(\frac{1}{%
4\alpha ^{2}}+\frac{2}{3}\beta ^{2}).
\end{equation}
where $\alpha =J\lambda t>1$ and $\beta =J(\lambda t)^{2}\ll 1$.
In particular, the  $t_{\min }$ via minimizing  
$(\Delta \hat{J}_{\vec{n}_{\perp }}^{2})_{\min}$ can be obtained by solving
\begin{equation}
\ \frac{d(\Delta \hat{J}_{\vec{n}_{\perp }}^{2})_{\min}}{dt}|_{t_{\min}}=0,
\end{equation}
which results in the optimal squeezing time
\begin{equation}
t_{\min}\simeq \frac{3^{1/6}}{\lambda }(2J)^{-2/3}.
\end{equation}
Inserting $t_{\min}$ into Eqs. (11) and (13), we now obtain the optimal squeezing
parameter
\begin{equation}
\xi _{opt}^{2}\simeq \frac{1}{2}(\frac{2J}{3})^{-2/3}.
\end{equation}
It is observed that the degree of spin squeezing  enhanced exponentially
by increasing the spin number $N$. 

\subsection{Decoherence effects}

Since any quantum system would suffer from decoherence, we now turn to considering
the dynamics of the system in the realistic case.
As $T_{1}$ for SiV centers can reach $1$ second in the low-temperature environment,
the spin relaxation can be ignored in our discussion.
Taking the dephasing ($\gamma _{s}$)
of the SiV spins and the decay of the mechanical modes ($\gamma _{m}$)
into consideration, the full dynamics of this system can be described by the following
master equation:
\begin{eqnarray}
\frac{d\hat{\rho}(t)}{dt} &=&-i[\hat{H}_{eff},\hat{\rho}]+\gamma _{s}%
\mathcal{D}[\hat{J}_{z}]\rho   \nonumber \\
&&+(n_{th}+1)\Gamma _{m}\mathcal{D}(\hat{J}_{-})+n_{th}\Gamma _{m}\mathcal{D}%
(\hat{J}_{+}),
\end{eqnarray}
where $n_{th}=(e^{\hslash  \omega _{m}/k_{B}T}-1)^{-1}$ is
the thermal phonon number at the environment temperature $T$
and $\mathcal{D}(\hat{o})\hat{\rho}=\hat{o}\hat{\rho}\hat{o}^{\dagger }-\frac{1}{%
2}\hat{o}^{\dagger }\hat{o}\hat{\rho}-\frac{1}{2}\hat{\rho}\hat{o}^{\dagger }\hat{o}$
for a given operator $\hat{o}$.
 The last two terms describe collective spin relaxation induced by mechanical dissipation,
 with $\Gamma _{m}=\gamma _{m}g_{e}^{2}/\Delta ^{2}$.

 \begin{figure}[t]
\includegraphics[width=8.4cm]{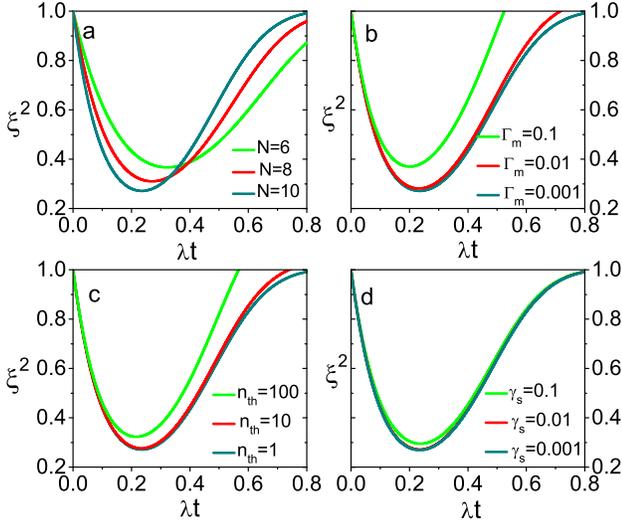}
\caption{\label{fig_2}  The effects of spin number and
decay parameters to  spin squeezing. 
The dynamic evolution of the system is simulated with different:
 (a) spin number $N$, where $\Gamma _{m}=0.001\lambda $, 
  $n_{th}=1$, and $\gamma _{s}=0.01\lambda $; (b) effective mechanical
decay rate $\Gamma _{m}$, where $N=10$, 
 $n_{th}=1$, and $\gamma _{s}=0.01\lambda $; 
 (c) thermal phonon number $n_{th}$, where $N=10$, 
 $\Gamma _{m}=0.001\lambda $, and $\gamma _{s}=0.01\lambda $;
(d) spin dephasing rate $\gamma _{s}$, where $N=10$, $\Gamma
_{m}=0.001\lambda $, and $n_{th}=1$.
}
\end{figure}

To study the effects of decoherence processes  to 
spin squeezing, we perform numerical simulations by solving the  master Eq. (17)
with effective Hamiltonian (10). The simulations are limited
in the case for small spin number $N$, and
the  initial state is set as the coherent spin state $|\psi _{0}\rangle $ along the $x$ axis.
As illustrated in Fig~2(a), we observe that the spin
squeezing can be enhanced by increasing the spin number.
In particular, with the increase in $N$, the time that
the system needs to reach the  optimal squeezing point will
be shortened. In Figs~2(b) and 2(c),
we show that the decoherence  of the mechanical mode 
has an obviously harmful effect to the squeezing dynamics.
In practical situations, this effect depends on not only the effective mechanical decay rate $\Gamma _{m}$,
but also the thermal phonon number $n_{th}$.
The dynamic evolution of the system with different spin dephasing rates is displayed in
 Fig~2(d). According to the simulation results, the effect of spin dephasing
 is not so obviously,  compared with the mechanical decay.

\begin{figure}[b]
\includegraphics[width=8.4cm]{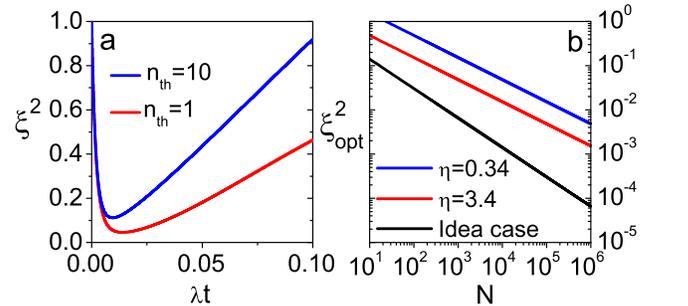}
\caption{\label{fig_3} (a) The spin squeezing dynamics
in the practical situation with spin number $N=1000$. The related decoherence parameters
are $\Gamma _{m}=0.001\lambda $, $\gamma _{s}=0.01\lambda $, and $n_{th}=\{1,10\}$.
(b) Optimal squeezing parameter $\xi _{opt}^{2}$ versus spin number $N$ in the ideal and realistic cases.
 In the practical case, the single-spin coupling-to-decay ratios are $\eta =\{3.4,0.34\}$ for  $n_{th}=\{1,10\}$, respectively.
}
\end{figure}

For spin-mechanical hybrid system with large spin number N,
directly solving the master Eq. (17) turn out to be difficult.
However, we can still study the dynamics of the system by estimating the mean values of the spin operators.
As  shown in Eq. (12), in order to estimate the optimal squeezing parameter in our system,
 a set of spin operators
$\{\langle \hat{J}_{x}\rangle ,\langle \hat{J}_{y}\rangle ,\langle \hat{J}%
_{z}\rangle ,\langle \hat{J}_{y}^{2}\rangle ,\langle \hat{J}_{z}^{2}\rangle
,\langle \hat{J}_{yz}\rangle \}$ need to be determined, where $\hat{J}_{yz}=\frac{1}{2}\langle \hat{%
J}_{y}\hat{J}_{z}+\hat{J}_{z}\hat{J}_{y}\rangle $.
Because the initial state satisfying $\hat{J}_{x}|\psi _{0}\rangle =J|\psi _{0}\rangle$,
we have ${\langle \hat{J}_{y}\rangle =\langle \hat{J}_{z}\rangle =\langle
\hat{J}_{yz}\rangle =0}$ and ${\langle \hat{J}_{y}^{2}\rangle =\langle \hat{J%
}_{z}^{2}\rangle =J/2}$
at the beginning.
As theoretically studied in Ref. \cite{PhysRevLett.110.156402},
the mechanical decay ($\Gamma _{m}$)
and  spin dephasing ($\gamma _{s}$)
terms can be  treated separately using an approximate numerical approach, and
the master Eq. (17) can be linearized in the short time regime.
As a result, the evolution of the spin averages can be described by
\begin{eqnarray}
\frac{d\langle \hat{J}_{x}\rangle }{dt} &=&-\gamma _{s}\langle \hat{J}%
_{x}\rangle   \nonumber \\
\frac{d\langle \hat{J}_{y}\rangle }{dt} &=&\lambda J\langle \hat{J}%
_{z}\rangle -\gamma _{s}\langle \hat{J}_{y}\rangle -\Gamma _{m}(\tilde{n}%
_{th}+\frac{1}{2})\langle \hat{J}_{y}\rangle +\Gamma _{m}\langle \hat{J}%
_{yz}\rangle   \nonumber \\
\frac{d\langle \hat{J}_{z}\rangle }{dt} &=&-2\Gamma _{m}(\tilde{n}_{th}+%
\frac{1}{2})\langle \hat{J}_{z}\rangle -\Gamma _{m}[J(J+1)-\langle \hat{J}%
_{z}^{2}\rangle ]  \nonumber \\
\frac{d\langle \hat{J}_{y}^{2}\rangle }{dt} &=&2J\lambda \langle \hat{J}%
_{yz}\rangle -2\gamma _{s}(\langle \hat{J}_{y}^{2}\rangle -\frac{J}{2}%
)+\Gamma _{m}(J+\frac{1}{2})\langle \hat{J}_{z}\rangle   \nonumber \\
&&-2\Gamma _{m}(\tilde{n}_{th}+\frac{1}{2})\langle \hat{J}_{y}^{2}-\hat{J}%
_{z}^{2}\rangle   \nonumber \\
\frac{d\langle \hat{J}_{z}^{2}\rangle }{dt} &=&-2\Gamma _{m}(\tilde{n}_{th}+%
\frac{1}{2})[3\langle \hat{J}_{z}^{2}\rangle -J(J+1)]  \nonumber \\
&&+\Gamma _{m}\langle \hat{J}_{z}\rangle \lbrack 1-2J(J+\frac{1}{2})]
\nonumber \\
\frac{d\langle \hat{J}_{yz}\rangle }{dt} &=&\lambda J\langle \hat{J}%
_{z}^{2}\rangle -\gamma _{s}\langle \hat{J}_{yz}\rangle -5(\tilde{n}_{th}+%
\frac{1}{2})\Gamma _{m}\langle \hat{J}_{yz}\rangle   \nonumber \\
&&-\Gamma _{m}(J^{2}-\frac{1}{4})\langle \hat{J}_{y}\rangle .
\end{eqnarray}
Putting Eqs. (11), (12), and (18) together,
we can study the  dynamic evolution of our system 
by numerical simulations.
The result is shown in Fig 3(a).
As discussed in Sec. \uppercase\expandafter{\romannumeral3},
the relative coupling (decay) parameters are $N\sim 10^{3}$,
 $\lambda \sim 10$ MHz,
$\Gamma _{m}\simeq \gamma _{m}g_{e}^{2}/\Delta ^{2}\sim 10$
 kHz, and $\gamma _{s} \sim 100$ kHz.
 As the system evolves, we show that the optimal squeezing parameters $\{0.046,0.112\}$ can be reached for
 $n_{th}=\{1,10\}$, respectively.

As illustrated in Ref. \cite{PhysRevLett.110.156402} for NV centers, the linearization
of the master equation may also result in some formulas to estimate the
optimal squeezing parameter for a system with arbitrary large spin number $N$.
By introducing  the single-spin coupling-to-decay ratio $\eta =g /\max\{n_{th}\gamma _{m},\gamma _{s}\}$,
the optimal squeezing parameter
can be estimated as
\begin{equation}
\text{\ }\xi _{opt}^{2}\simeq \frac{2}{\sqrt{J\eta }},
\end{equation}
with optimal squeezing time $t_{\min}\simeq 1/(\gamma _{s}\sqrt{J\eta })$.
In comparison, the  estimated optimal spin squeezing parameters in the ideal case
and in the realistic case are shown in Fig~3 (b).
These results are obtained by solving the Eqs. (16) and (19), respectively.
We observe that for both the two cases, the spin  squeezing degree can be enhanced
exponentially by increasing the spin number.

\section{Discussion and Conclusion}

We now take a further consideration about the phonon number  fluctuations in our system. 　
In particular, we mention that the phonon fluctuation term
$\propto \hat{b}^{\dagger }\hat{b}\hat{J}_{z}$
in effective Hamiltonian (10) could also be utilized
for spin squeezing through a cavity feedback scheme.
Remarkably,  this cavity feedback spin squeezing protocol has already been
experimentally studied with ultracold atoms
\cite{PhysRevLett.94.023003,PhysRevLett.104.073602,PhysRevA.81.021804}.
In our setup, it can be performed 　
by driving the mechanical phonon modes \cite{PhysRevLett.110.156402}.
On the other hand, the phonon number $n=\hat{b}^{\dagger }\hat{b}$ couples
to $\hat{J}_{z}$, which may also result in additional spin
dephasing due to thermal number fluctuations.
Together with the inhomogeneous broadening of the spin ensemble, it may
limit the long spin-coherence time of SiV centers.
However, we remind that these effects can be effectively
suppressed by  the spin-echo techniques
\cite{PhysRevLett.110.250503,PhysRevX.5.031031}. 

In summary, we present an efficient scheme for squeezing the
SiV center spins in an exquisite spin-mechanical hybrid device. 
We show that, the strain couplings between the SiV spins and the
compression phonon modes allow long-range spin-spin interactions in the large detuning condition.   
By  modeling these  nonlinear spin-spin couplings with an effective one-axis twisting Hamiltonian,　
we find that the SiV centers  with properly initial states 
can be steered to the SSSs as the system evolves.
Further more, we study the  effects of decoherence processes 
 with numerical simulations,
 showing that in the practical conditions, the SiV spin ensemble with a spin number $N\sim 10^{3}$,  
 can be strongly squeezed in the low-temperature environment $T\sim 100$ mK .
This spin-mechanical hybrid architecture provides a promising platform
for studying the SSSs with spin-phonon couplings,
which may further be  applied in entanglement detection,
 quantum information, and high-precision measurement.

\section*{Acknowledgments}

Bo Li thanks  Yuan Zhou for fruitful discussions.
Pengbo Li was supported
by the NSFC under Grant Nos. 11774285,
91536115 and 11534008, and the Fundamental Research
Funds for the Central Universities.
Tongcang Li acknowledges 
supports from the Gordon and Betty Moore Foundation, the Purdue EFC research grant, and the NSF under grant No. PHY-1555035.
Part of the simulations are coded in PYTHON using the QUTIP library \cite{CPC}.

\appendix

\section{Calculation of mechanical compression modes in diamond waveguide}

In the main text we show that our setup is realized by a diamond waveguide with size $(l,w,t)$,
and we focus on the case $l\gg \{w,t\}$. 
We now discuss the mechanical compression modes propagating along the waveguide.
With the framework of elastic mechanical theory,
the  mechanical modes can be treated as a continuous displacement field $\vec{u}(\vec{r},t)$. 
 For a linear, isotropic medium, the motional equation
of the time-dependent displacement field  reads
\begin{equation}
\rho \frac{\partial ^{2}}{\partial t^{2}}\vec{u}(\vec{r},t)=(\lambda +\mu )%
\vec{\nabla}(\vec{\nabla}\cdot \vec{u}(\vec{r},t))+\mu \vec{\nabla}^{2}\vec{u%
}(\vec{r},t),
\end{equation}
where $\rho $ is the mass density of the diamond. Specifically, the Lam{\'e} constants $\lambda $ and $\mu $ can
be expressed in terms of the Young's modulus $E$ and the Poisson ratio $\nu $, i.e., 
\begin{equation}
\lambda =\frac{\nu E}{(1+\nu )(1-2\nu )},\mu =\frac{E}{2(1+\nu )}.
\end{equation}%
For a diamond waveguide, we have $\rho =3500$ kg/m$^{3}$, $E=1050$ Gpa and $\nu =0.2$.
By assuming periodic boundary conditions, the motional equation (A1) can be solved by the general ansatz
\begin{equation}
\vec{u}(\vec{r},t)=\frac{1}{\sqrt{2}}\sum\limits_{k,n}\vec{u}_{n,k}^{\perp
}(y,z)[A_{n,k}(t)e^{ikx}+A_{n,k}^{\ast }e^{-ikx}],
\end{equation}
where  the index $n$ denotes the different phonon branches, and
$k=2\pi /L\times m$ labels the wavevector along the $x$-axis.
Note that the amplitudes $A_{n,k}(t)$ and the mode frequencies $\omega _{n,k}$
satisfy   $\ddot{A}_{n,k}(t)+\omega _{n,k}^{2}A_{n,k}(t)=0$. Together with 
the transverse mode profile $\vec{u}_{n,k}^{\perp }(x,y)$, they can be
calculated by the finite-element-method simulations for a waveguide
with arbitrary size.

\section{Quantization of the displacement field}

We now consider the  propagating phonon modes 
supported by the diamond waveguide in the quantum regime.
Using a quantization method similar to that used for
the electromagnetic field in quantum optics, the
motion of the diamond lattices  associated with the  propagating phonons
can be quantized. As a result,
the quantized  displacement field 
 can be represented as
 \begin{equation}
\vec{u}(\vec{r})=\sum\limits_{k,n}Q_{n,k}^{0}\vec{u}_{n,k}^{\perp }(y,z)(%
\hat{b}_{n,k}e^{ikx}+\hat{b}_{n,k}^{\dagger }e^{-ikx}),
\end{equation}
where $\hat{b}_{n,k}$ and $\hat{b}_{n,k}^{\dagger }$  are the annihilation and creation operators
of the phonons, 
 which can be defined as
\begin{equation}
\hat{b}_{n,-k}=\frac{1}{2}(\frac{\hat{Q}_{n,k}}{Q_{n,k}^{0}}+\frac{i\hat{P}%
_{n,k}}{P_{n,k}^{0}}),\hat{b}_{n,k}^{\dagger }=\frac{1}{2}(\frac{\hat{Q}%
_{n,k}}{Q_{n,k}^{0}}-\frac{i\hat{P}_{n,k}}{P_{n,k}^{0}}).
\end{equation}
Here $\hat{Q}_{n,k}$ and  $\hat{P}_{n,k}$ denote the generalized coordinate and momentum operators,
which satisfying the canonical commutation relations
$\lbrack \hat{Q}_{n,k},\hat{P}_{n,k}]=i\hslash \delta _{n,n^{\prime
}}\delta _{kk^{\prime }}$.
In particular, the effective zero point
fluctuations of the phonon modes can be estimated as
 $Q_{n,k}^{0}=\sqrt{\hslash /(2\rho V_{eff}\omega _{n,k})}$ and $P%
_{n,k}^{0}=\sqrt{(\hslash \rho V_{eff}\omega _{n,k})/2}$. Further more,
the resulting Hamiltonian for the  propagating phonons reads
\begin{equation}
\hat{H}_{ph}=\sum\limits_{k,n}\hbar \omega _{n,k}\hat{b}_{n,k}^{\dagger }%
\hat{b}_{n,k}.
\end{equation}

\section{Description of the spin-phonon couplings}

In our system, the time-dependent displacements  of the
lattice atoms forming the defect centers affect the defects' electronic structure, 
which results in the strain couplings between the
propagating phonons and the orbital degrees of freedom of the SiV centers. 
As described in the main text,
we assume the symmetry axis of the SiV centers is along the $z$ axis,
and the propagating phonon modes  supported by the waveguide  is  along the $x$ axis.
Then the strain tensor  can be defined as 
\begin{equation}
\epsilon _{ij}=\frac{1}{2}(\frac{\partial u_{i}}{\partial x_{j}}+\frac{%
\partial u_{j}}{\partial x_{i}}),
\end{equation}
with $u_{i(j)}$ denoting the quantized displacement field with subscripts $i,j\in \{x,y,z\}$.
 For a small stress, the effect of lattice deformation is linear in the
strain components and is captured by Hamiltonian
\begin{equation}
\hat{H}_{strain}=\sum\limits_{i,j}V_{ij}\epsilon _{ij},
\end{equation}
where $V_{ij}$ are  the particular strain components  acting on the
SiV electronic levels.

The Hamiltonian $H_{strain}$ can be rewritten with
Group theory in terms of basis-independent linear combinations
of strain components adapted to the symmetries of the SiV center.
These combinations can be viewed as particular modes of deformation,
and the resulting effects on the orbital wave functions can be
deduced using group theory. 
Specifically, by projecting the strain tensor
onto the irreducible representations of the point group of the SiV center ($D_{3d}$),
 we arrive at
\begin{equation}
\hat{H}_{strain}=\sum\nolimits_{l}V_{l}\epsilon _{l}, 
\end{equation}
where each $\epsilon _{l}$ denotes a linear combination of strain components $\epsilon _{ij}$,
and corresponds to specific symmetries indicated by the
subscript $l$, i.e.,
\begin{eqnarray}
\epsilon _{A_{1g}} &=&t_{\perp }(\epsilon _{xx}+\epsilon _{yy})+t_{\parallel
}\epsilon _{zz},  \nonumber \\
\epsilon _{E_{gx}} &=&d(\epsilon _{xx}-\epsilon _{yy})+f\epsilon _{zx},  \\
\epsilon _{E_{gy}} &=&-2d\epsilon _{xy}+f\epsilon _{yz}. \nonumber
\end{eqnarray}
Here, $t_{\perp },t_{\parallel },d$ and $f$ are four
strain-susceptibility parameters, which
are related to the original stress-response coefficients. 
 The effects of these strain components on the electronic states can be represented as
\begin{eqnarray}
V_{A_{1g}} &=&|e_{x}\rangle \langle e_{x}|+|e_{y}\rangle \langle e_{y}|, \nonumber \\
V_{E_{gx}} &=&|e_{x}\rangle \langle e_{x}|-|e_{y}\rangle \langle e_{y}|,  \\
V_{E_{gy}} &=&|e_{x}\rangle \langle e_{y}|+|e_{y}\rangle \langle e_{x}|. \nonumber
\end{eqnarray}

As coupling to symmetric local distortions $(\sim \epsilon _{A_{1g}})$ shifts all ground states equally,
it can be ignored in our discussion.
In the basis spanned by the eigenstates of  the spin-orbit coupling
$\left\{ |e_{-}\rangle ,|e_{+}\rangle \right\} $, the strain Hamiltonian
can be rewritten as
\begin{equation}
\hat{H}_{strain}=\epsilon _{E_{gx}}(\hat{L}_{-}+\hat{L}_{+})-i\epsilon _{E_{gy}}(%
\hat{L}_{-}-\hat{L}_{+}),
\end{equation}
where $\hat{L}_{+}=(\hat{L}_{-})^{\dagger }=|c\rangle \langle
a|+|d\rangle \langle b|$ is the
orbital raising operator within the ground states.
By decomposing the local displacement
field as in Eq. (B1),
the spin-phonon couplings
under the rotating-wave approximation can be  expressed as
\begin{equation}
\hat{H}_{strain}\simeq \sum\limits_{j,n,k}[(g_{n,k}^{j}\hat{\sigma}%
_{+\uparrow }^{j}+g_{n,-k}^{j\ast }\hat{\sigma}_{+\downarrow }^{j})\hat{b}%
_{n,k}e^{ikx_{j}}+H.c.],
\end{equation} 
where $\hat{\sigma}_{+\uparrow }^{j}=|c^{j}\rangle \langle a^{j}|$ and $%
\hat{\sigma}_{+\downarrow }^{j}=|d^{j}\rangle \langle b^{j}|$. 
Here the spin-phonon coupling strength is given by
\begin{equation}
g_{n,k}=\frac{d}{v_{l}}\sqrt{\frac{\hbar \omega _{n,k}}{2\rho lwt}}\zeta
_{n,k}(\vec{r}),
\end{equation}
and the dimensionless profile reads  
\begin{eqnarray}
\zeta _{n,k}(\vec{r}) &=&\frac{1}{|k|}[(iku_{n,k}^{\perp ,x}+ik\frac{f}{2d}%
u_{n,k}^{\perp ,z}+\frac{f}{2d}\partial _{z}u_{n,k}^{\perp ,x}-\partial
_{y}u_{n,k}^{\perp ,y})  \nonumber \\
&&-i(iku_{n,k}^{\perp ,y}+\partial _{y}u_{n,k}^{\perp ,x}+\frac{f}{2d}%
\partial _{y}u_{n,k}^{\perp ,z}+\frac{\partial _{z}u_{n,k}^{\perp ,y}}{2})]. \nonumber \\
\end{eqnarray} 
For the compression phonon modes propagating along the $x$-axis, 
the mode function can be approximated as 
 $\vec{u}_{n,k}(\vec{r})\sim \vec{e}_{x}\cos(\omega_{n,k}x/v_{l})$, resulting in 
$|\zeta _{n,k}(\vec{r})|\sim 1$.


\begin{thebibliography}{60}%
\makeatletter
\providecommand \@ifxundefined [1]{%
 \@ifx{#1\undefined}
}%
\providecommand \@ifnum [1]{%
 \ifnum #1\expandafter \@firstoftwo
 \else \expandafter \@secondoftwo
 \fi
}%
\providecommand \@ifx [1]{%
 \ifx #1\expandafter \@firstoftwo
 \else \expandafter \@secondoftwo
 \fi
}%
\providecommand \natexlab [1]{#1}%
\providecommand \enquote  [1]{``#1''}%
\providecommand \bibnamefont  [1]{#1}%
\providecommand \bibfnamefont [1]{#1}%
\providecommand \citenamefont [1]{#1}%
\providecommand \href@noop [0]{\@secondoftwo}%
\providecommand \href [0]{\begingroup \@sanitize@url \@href}%
\providecommand \@href[1]{\@@startlink{#1}\@@href}%
\providecommand \@@href[1]{\endgroup#1\@@endlink}%
\providecommand \@sanitize@url [0]{\catcode `\\12\catcode `\$12\catcode
  `\&12\catcode `\#12\catcode `\^12\catcode `\_12\catcode `\%12\relax}%
\providecommand \@@startlink[1]{}%
\providecommand \@@endlink[0]{}%
\providecommand \url  [0]{\begingroup\@sanitize@url \@url }%
\providecommand \@url [1]{\endgroup\@href {#1}{\urlprefix }}%
\providecommand \urlprefix  [0]{URL }%
\providecommand \Eprint [0]{\href }%
\providecommand \doibase [0]{https://doi.org/}%
\providecommand \selectlanguage [0]{\@gobble}%
\providecommand \bibinfo  [0]{\@secondoftwo}%
\providecommand \bibfield  [0]{\@secondoftwo}%
\providecommand \translation [1]{[#1]}%
\providecommand \BibitemOpen [0]{}%
\providecommand \bibitemStop [0]{}%
\providecommand \bibitemNoStop [0]{.\EOS\space}%
\providecommand \EOS [0]{\spacefactor3000\relax}%
\providecommand \BibitemShut  [1]{\csname bibitem#1\endcsname}%
\let\auto@bib@innerbib\@empty
\bibitem [{\citenamefont {Doherty}\ \emph {et~al.}(2013)\citenamefont
  {Doherty}, \citenamefont {Manson}, \citenamefont {Delaney}, \citenamefont
  {Jelezko}, \citenamefont {Wrachtrup},\ and\ \citenamefont
  {Hollenberg}}]{PR-528-1}%
  \BibitemOpen
  \bibfield  {author} {\bibinfo {author} {\bibfnamefont {M.~W.}\ \bibnamefont
  {Doherty}}, \bibinfo {author} {\bibfnamefont {N.~B.}\ \bibnamefont {Manson}},
  \bibinfo {author} {\bibfnamefont {P.}~\bibnamefont {Delaney}}, \bibinfo
  {author} {\bibfnamefont {F.}~\bibnamefont {Jelezko}}, \bibinfo {author}
  {\bibfnamefont {J.}~\bibnamefont {Wrachtrup}},\ and\ \bibinfo {author}
  {\bibfnamefont {L.~C.~L.}\ \bibnamefont {Hollenberg}},\ }\bibfield  {title}
  {\bibinfo {title} {The nitrogen-vacancy colour centre in diamond},\ }\href
  {https://doi.org/https://doi.org/10.1016/j.physrep.2013.02.001} {\bibfield
  {journal} {\bibinfo  {journal} {Phys. Rep.}\ }\textbf {\bibinfo {volume}
  {528}},\ \bibinfo {pages} {1 } (\bibinfo {year} {2013})}\BibitemShut
  {NoStop}%
\bibitem [{\citenamefont {Hepp}\ \emph {et~al.}(2014)\citenamefont {Hepp},
  \citenamefont {M\"uller}, \citenamefont {Waselowski}, \citenamefont {Becker},
  \citenamefont {Pingault}, \citenamefont {Sternschulte}, \citenamefont
  {Steinm\"uller-Nethl}, \citenamefont {Gali}, \citenamefont {Maze},
  \citenamefont {Atat\"ure},\ and\ \citenamefont
  {Becher}}]{PhysRevLett.112.036405}%
  \BibitemOpen
  \bibfield  {author} {\bibinfo {author} {\bibfnamefont {C.}~\bibnamefont
  {Hepp}}, \bibinfo {author} {\bibfnamefont {T.}~\bibnamefont {M\"uller}},
  \bibinfo {author} {\bibfnamefont {V.}~\bibnamefont {Waselowski}}, \bibinfo
  {author} {\bibfnamefont {J.~N.}\ \bibnamefont {Becker}}, \bibinfo {author}
  {\bibfnamefont {B.}~\bibnamefont {Pingault}}, \bibinfo {author}
  {\bibfnamefont {H.}~\bibnamefont {Sternschulte}}, \bibinfo {author}
  {\bibfnamefont {D.}~\bibnamefont {Steinm\"uller-Nethl}}, \bibinfo {author}
  {\bibfnamefont {A.}~\bibnamefont {Gali}}, \bibinfo {author} {\bibfnamefont
  {J.~R.}\ \bibnamefont {Maze}}, \bibinfo {author} {\bibfnamefont
  {M.}~\bibnamefont {Atat\"ure}},\ and\ \bibinfo {author} {\bibfnamefont
  {C.}~\bibnamefont {Becher}},\ }\bibfield  {title} {\bibinfo {title}
  {Electronic structure of the silicon vacancy color center in diamond},\
  }\href {https://doi.org/10.1103/PhysRevLett.112.036405} {\bibfield  {journal}
  {\bibinfo  {journal} {Phys. Rev. Lett.}\ }\textbf {\bibinfo {volume} {112}},\
  \bibinfo {pages} {036405} (\bibinfo {year} {2014})}\BibitemShut {NoStop}%
\bibitem [{\citenamefont {Sipahigil}\ \emph {et~al.}(2014)\citenamefont
  {Sipahigil}, \citenamefont {Jahnke}, \citenamefont {Rogers}, \citenamefont
  {Teraji}, \citenamefont {Isoya}, \citenamefont {Zibrov}, \citenamefont
  {Jelezko},\ and\ \citenamefont {Lukin}}]{PhysRevLett.113.113602}%
  \BibitemOpen
  \bibfield  {author} {\bibinfo {author} {\bibfnamefont {A.}~\bibnamefont
  {Sipahigil}}, \bibinfo {author} {\bibfnamefont {K.~D.}\ \bibnamefont
  {Jahnke}}, \bibinfo {author} {\bibfnamefont {L.~J.}\ \bibnamefont {Rogers}},
  \bibinfo {author} {\bibfnamefont {T.}~\bibnamefont {Teraji}}, \bibinfo
  {author} {\bibfnamefont {J.}~\bibnamefont {Isoya}}, \bibinfo {author}
  {\bibfnamefont {A.~S.}\ \bibnamefont {Zibrov}}, \bibinfo {author}
  {\bibfnamefont {F.}~\bibnamefont {Jelezko}},\ and\ \bibinfo {author}
  {\bibfnamefont {M.~D.}\ \bibnamefont {Lukin}},\ }\bibfield  {title} {\bibinfo
  {title} {Indistinguishable photons from separated silicon-vacancy centers in
  diamond},\ }\href {https://doi.org/10.1103/PhysRevLett.113.113602} {\bibfield
   {journal} {\bibinfo  {journal} {Phys. Rev. Lett.}\ }\textbf {\bibinfo
  {volume} {113}},\ \bibinfo {pages} {113602} (\bibinfo {year}
  {2014})}\BibitemShut {NoStop}%
\bibitem [{\citenamefont {Sipahigil}\ \emph {et~al.}(2016)\citenamefont
  {Sipahigil}, \citenamefont {Evans}, \citenamefont {Sukachev}, \citenamefont
  {Burek}, \citenamefont {Borregaard}, \citenamefont {Bhaskar}, \citenamefont
  {Nguyen}, \citenamefont {Pacheco}, \citenamefont {Atikian}, \citenamefont
  {Meuwly}, \citenamefont {Camacho}, \citenamefont {Jelezko}, \citenamefont
  {Bielejec}, \citenamefont {Park}, \citenamefont {Lon{\v c}ar},\ and\
  \citenamefont {Lukin}}]{Sipahigil847}%
  \BibitemOpen
  \bibfield  {author} {\bibinfo {author} {\bibfnamefont {A.}~\bibnamefont
  {Sipahigil}}, \bibinfo {author} {\bibfnamefont {R.~E.}\ \bibnamefont
  {Evans}}, \bibinfo {author} {\bibfnamefont {D.~D.}\ \bibnamefont {Sukachev}},
  \bibinfo {author} {\bibfnamefont {M.~J.}\ \bibnamefont {Burek}}, \bibinfo
  {author} {\bibfnamefont {J.}~\bibnamefont {Borregaard}}, \bibinfo {author}
  {\bibfnamefont {M.~K.}\ \bibnamefont {Bhaskar}}, \bibinfo {author}
  {\bibfnamefont {C.~T.}\ \bibnamefont {Nguyen}}, \bibinfo {author}
  {\bibfnamefont {J.~L.}\ \bibnamefont {Pacheco}}, \bibinfo {author}
  {\bibfnamefont {H.~A.}\ \bibnamefont {Atikian}}, \bibinfo {author}
  {\bibfnamefont {C.}~\bibnamefont {Meuwly}}, \bibinfo {author} {\bibfnamefont
  {R.~M.}\ \bibnamefont {Camacho}}, \bibinfo {author} {\bibfnamefont
  {F.}~\bibnamefont {Jelezko}}, \bibinfo {author} {\bibfnamefont
  {E.}~\bibnamefont {Bielejec}}, \bibinfo {author} {\bibfnamefont
  {H.}~\bibnamefont {Park}}, \bibinfo {author} {\bibfnamefont {M.}~\bibnamefont
  {Lon{\v c}ar}},\ and\ \bibinfo {author} {\bibfnamefont {M.~D.}\ \bibnamefont
  {Lukin}},\ }\bibfield  {title} {\bibinfo {title} {An integrated diamond
  nanophotonics platform for quantum-optical networks},\ }\href
  {https://doi.org/10.1126/science.aah6875} {\bibfield  {journal} {\bibinfo
  {journal} {Science}\ }\textbf {\bibinfo {volume} {354}},\ \bibinfo {pages}
  {847} (\bibinfo {year} {2016})}\BibitemShut {NoStop}%
\bibitem [{\citenamefont {Bhaskar}\ \emph {et~al.}(2017)\citenamefont
  {Bhaskar}, \citenamefont {Sukachev}, \citenamefont {Sipahigil}, \citenamefont
  {Evans}, \citenamefont {Burek}, \citenamefont {Nguyen}, \citenamefont
  {Rogers}, \citenamefont {Siyushev}, \citenamefont {Metsch}, \citenamefont
  {Park}, \citenamefont {Jelezko}, \citenamefont {Lon\ifmmode~\check{c}\else
  \v{c}\fi{}ar},\ and\ \citenamefont {Lukin}}]{PhysRevLett.118.223603}%
  \BibitemOpen
  \bibfield  {author} {\bibinfo {author} {\bibfnamefont {M.~K.}\ \bibnamefont
  {Bhaskar}}, \bibinfo {author} {\bibfnamefont {D.~D.}\ \bibnamefont
  {Sukachev}}, \bibinfo {author} {\bibfnamefont {A.}~\bibnamefont {Sipahigil}},
  \bibinfo {author} {\bibfnamefont {R.~E.}\ \bibnamefont {Evans}}, \bibinfo
  {author} {\bibfnamefont {M.~J.}\ \bibnamefont {Burek}}, \bibinfo {author}
  {\bibfnamefont {C.~T.}\ \bibnamefont {Nguyen}}, \bibinfo {author}
  {\bibfnamefont {L.~J.}\ \bibnamefont {Rogers}}, \bibinfo {author}
  {\bibfnamefont {P.}~\bibnamefont {Siyushev}}, \bibinfo {author}
  {\bibfnamefont {M.~H.}\ \bibnamefont {Metsch}}, \bibinfo {author}
  {\bibfnamefont {H.}~\bibnamefont {Park}}, \bibinfo {author} {\bibfnamefont
  {F.}~\bibnamefont {Jelezko}}, \bibinfo {author} {\bibfnamefont
  {M.}~\bibnamefont {Lon\ifmmode~\check{c}\else \v{c}\fi{}ar}},\ and\ \bibinfo
  {author} {\bibfnamefont {M.~D.}\ \bibnamefont {Lukin}},\ }\bibfield  {title}
  {\bibinfo {title} {Quantum nonlinear optics with a germanium-vacancy color
  center in a nanoscale diamond waveguide},\ }\href
  {https://doi.org/10.1103/PhysRevLett.118.223603} {\bibfield  {journal}
  {\bibinfo  {journal} {Phys. Rev. Lett.}\ }\textbf {\bibinfo {volume} {118}},\
  \bibinfo {pages} {223603} (\bibinfo {year} {2017})}\BibitemShut {NoStop}%
\bibitem [{\citenamefont {{Thiel}}\ \emph {et~al.}(2019)\citenamefont
  {{Thiel}}, \citenamefont {{Wang}}, \citenamefont {{Tschudin}}, \citenamefont
  {{Rohner}}, \citenamefont {{Guti{\'e}rrez-Lezama}}, \citenamefont {{Ubrig}},
  \citenamefont {{Gibertini}}, \citenamefont {{Giannini}}, \citenamefont
  {{Morpurgo}},\ and\ \citenamefont {{Maletinsky}}}]{Science-364-973}%
  \BibitemOpen
  \bibfield  {author} {\bibinfo {author} {\bibfnamefont {L.}~\bibnamefont
  {{Thiel}}}, \bibinfo {author} {\bibfnamefont {Z.}~\bibnamefont {{Wang}}},
  \bibinfo {author} {\bibfnamefont {M.~A.}\ \bibnamefont {{Tschudin}}},
  \bibinfo {author} {\bibfnamefont {D.}~\bibnamefont {{Rohner}}}, \bibinfo
  {author} {\bibfnamefont {I.}~\bibnamefont {{Guti{\'e}rrez-Lezama}}}, \bibinfo
  {author} {\bibfnamefont {N.}~\bibnamefont {{Ubrig}}}, \bibinfo {author}
  {\bibfnamefont {M.}~\bibnamefont {{Gibertini}}}, \bibinfo {author}
  {\bibfnamefont {E.}~\bibnamefont {{Giannini}}}, \bibinfo {author}
  {\bibfnamefont {A.~F.}\ \bibnamefont {{Morpurgo}}},\ and\ \bibinfo {author}
  {\bibfnamefont {P.}~\bibnamefont {{Maletinsky}}},\ }\bibfield  {title}
  {\bibinfo {title} {{Probing magnetism in 2D materials at the nanoscale with
  single-spin microscopy}},\ }\href {https://doi.org/10.1126/science.aav6926}
  {\bibfield  {journal} {\bibinfo  {journal} {Science}\ }\textbf {\bibinfo
  {volume} {364}},\ \bibinfo {pages} {973} (\bibinfo {year}
  {2019})}\BibitemShut {NoStop}%
\bibitem [{\citenamefont {Zu}\ \emph {et~al.}(2014)\citenamefont {Zu},
  \citenamefont {Wang}, \citenamefont {He}, \citenamefont {Zhang},
  \citenamefont {Dai}, \citenamefont {Wang},\ and\ \citenamefont
  {Duan}}]{Natrue-514-72}%
  \BibitemOpen
  \bibfield  {author} {\bibinfo {author} {\bibfnamefont {C.}~\bibnamefont
  {Zu}}, \bibinfo {author} {\bibfnamefont {W.-B.}\ \bibnamefont {Wang}},
  \bibinfo {author} {\bibfnamefont {L.}~\bibnamefont {He}}, \bibinfo {author}
  {\bibfnamefont {W.-G.}\ \bibnamefont {Zhang}}, \bibinfo {author}
  {\bibfnamefont {C.-Y.}\ \bibnamefont {Dai}}, \bibinfo {author} {\bibfnamefont
  {F.}~\bibnamefont {Wang}},\ and\ \bibinfo {author} {\bibfnamefont {L.-M.}\
  \bibnamefont {Duan}},\ }\bibfield  {title} {\bibinfo {title} {Experimental
  realization of universal geometric quantum gates with solid-state spins},\
  }\href {https://doi.org/10.1038/nature13729} {\bibfield  {journal} {\bibinfo
  {journal} {Nature (London)}\ }\textbf {\bibinfo {volume} {514}},\ \bibinfo
  {pages} {72} (\bibinfo {year} {2014})}\BibitemShut {NoStop}%
\bibitem [{\citenamefont {Chen}\ and\ \citenamefont
  {Yin}(2019)}]{PhysRevA.99.022319}%
  \BibitemOpen
  \bibfield  {author} {\bibinfo {author} {\bibfnamefont {X.-Y.}\ \bibnamefont
  {Chen}}\ and\ \bibinfo {author} {\bibfnamefont {Z.-Q.}\ \bibnamefont {Yin}},\
  }\bibfield  {title} {\bibinfo {title} {Universal quantum gates between
  nitrogen-vacancy centers in a levitated nanodiamond},\ }\href
  {https://doi.org/10.1103/PhysRevA.99.022319} {\bibfield  {journal} {\bibinfo
  {journal} {Phys. Rev. A}\ }\textbf {\bibinfo {volume} {99}},\ \bibinfo
  {pages} {022319} (\bibinfo {year} {2019})}\BibitemShut {NoStop}%
\bibitem [{\citenamefont {Xu}\ \emph {et~al.}(2019)\citenamefont {Xu},
  \citenamefont {Yin}, \citenamefont {Han},\ and\ \citenamefont
  {Li}}]{xu2019quantum}%
  \BibitemOpen
  \bibfield  {author} {\bibinfo {author} {\bibfnamefont {Z.}~\bibnamefont
  {Xu}}, \bibinfo {author} {\bibfnamefont {Z.-q.}\ \bibnamefont {Yin}},
  \bibinfo {author} {\bibfnamefont {Q.}~\bibnamefont {Han}},\ and\ \bibinfo
  {author} {\bibfnamefont {T.}~\bibnamefont {Li}},\ }\bibfield  {title}
  {\bibinfo {title} {Quantum information processing with closely-spaced diamond
  color centers in strain and magnetic fields},\ }\href@noop {} {\bibfield
  {journal} {\bibinfo  {journal} {arXiv preprint arXiv:1909.11775}\ } (\bibinfo
  {year} {2019})}\BibitemShut {NoStop}%
\bibitem [{\citenamefont {Yang}\ \emph {et~al.}(2011)\citenamefont {Yang},
  \citenamefont {Yin}, \citenamefont {Hu}, \citenamefont {Feng},\ and\
  \citenamefont {Du}}]{PhysRevA.84.010301}%
  \BibitemOpen
  \bibfield  {author} {\bibinfo {author} {\bibfnamefont {W.~L.}\ \bibnamefont
  {Yang}}, \bibinfo {author} {\bibfnamefont {Z.~Q.}\ \bibnamefont {Yin}},
  \bibinfo {author} {\bibfnamefont {Y.}~\bibnamefont {Hu}}, \bibinfo {author}
  {\bibfnamefont {M.}~\bibnamefont {Feng}},\ and\ \bibinfo {author}
  {\bibfnamefont {J.~F.}\ \bibnamefont {Du}},\ }\bibfield  {title} {\bibinfo
  {title} {High-fidelity quantum memory using nitrogen-vacancy center ensemble
  for hybrid quantum computation},\ }\href
  {https://doi.org/10.1103/PhysRevA.84.010301} {\bibfield  {journal} {\bibinfo
  {journal} {Phys. Rev. A}\ }\textbf {\bibinfo {volume} {84}},\ \bibinfo
  {pages} {010301} (\bibinfo {year} {2011})}\BibitemShut {NoStop}%
\bibitem [{\citenamefont {Li}\ \emph {et~al.}(2017)\citenamefont {Li},
  \citenamefont {Li}, \citenamefont {Zhou}, \citenamefont {Ma},\ and\
  \citenamefont {Li}}]{PhysRevA.96.032342}%
  \BibitemOpen
  \bibfield  {author} {\bibinfo {author} {\bibfnamefont {B.}~\bibnamefont
  {Li}}, \bibinfo {author} {\bibfnamefont {P.-B.}\ \bibnamefont {Li}}, \bibinfo
  {author} {\bibfnamefont {Y.}~\bibnamefont {Zhou}}, \bibinfo {author}
  {\bibfnamefont {S.-L.}\ \bibnamefont {Ma}},\ and\ \bibinfo {author}
  {\bibfnamefont {F.-L.}\ \bibnamefont {Li}},\ }\bibfield  {title} {\bibinfo
  {title} {Quantum microwave-optical interface with nitrogen-vacancy centers in
  diamond},\ }\href {https://doi.org/10.1103/PhysRevA.96.032342} {\bibfield
  {journal} {\bibinfo  {journal} {Phys. Rev. A}\ }\textbf {\bibinfo {volume}
  {96}},\ \bibinfo {pages} {032342} (\bibinfo {year} {2017})}\BibitemShut
  {NoStop}%
\bibitem [{\citenamefont {Xia}\ and\ \citenamefont
  {Twamley}(2015)}]{PhysRevA.91.042307}%
  \BibitemOpen
  \bibfield  {author} {\bibinfo {author} {\bibfnamefont {K.}~\bibnamefont
  {Xia}}\ and\ \bibinfo {author} {\bibfnamefont {J.}~\bibnamefont {Twamley}},\
  }\bibfield  {title} {\bibinfo {title} {Solid-state optical interconnect
  between distant superconducting quantum chips},\ }\href
  {https://doi.org/10.1103/PhysRevA.91.042307} {\bibfield  {journal} {\bibinfo
  {journal} {Phys. Rev. A}\ }\textbf {\bibinfo {volume} {91}},\ \bibinfo
  {pages} {042307} (\bibinfo {year} {2015})}\BibitemShut {NoStop}%
\bibitem [{\citenamefont {Li}\ \emph {et~al.}(2019)\citenamefont {Li},
  \citenamefont {Li}, \citenamefont {Zhou}, \citenamefont {Liu}, \citenamefont
  {Li},\ and\ \citenamefont {Li}}]{PhysRevApplied.11.044026}%
  \BibitemOpen
  \bibfield  {author} {\bibinfo {author} {\bibfnamefont {B.}~\bibnamefont
  {Li}}, \bibinfo {author} {\bibfnamefont {P.-B.}\ \bibnamefont {Li}}, \bibinfo
  {author} {\bibfnamefont {Y.}~\bibnamefont {Zhou}}, \bibinfo {author}
  {\bibfnamefont {J.}~\bibnamefont {Liu}}, \bibinfo {author} {\bibfnamefont
  {H.-R.}\ \bibnamefont {Li}},\ and\ \bibinfo {author} {\bibfnamefont {F.-L.}\
  \bibnamefont {Li}},\ }\bibfield  {title} {\bibinfo {title} {Interfacing a
  topological qubit with a spin qubit in a hybrid quantum system},\ }\href
  {https://doi.org/10.1103/PhysRevApplied.11.044026} {\bibfield  {journal}
  {\bibinfo  {journal} {Phys. Rev. Applied}\ }\textbf {\bibinfo {volume}
  {11}},\ \bibinfo {pages} {044026} (\bibinfo {year} {2019})}\BibitemShut
  {NoStop}%
\bibitem [{\citenamefont {Yin}\ \emph {et~al.}(2015)\citenamefont {Yin},
  \citenamefont {Zhao},\ and\ \citenamefont {Li}}]{yin2015hybrid}%
  \BibitemOpen
  \bibfield  {author} {\bibinfo {author} {\bibfnamefont {Z.}~\bibnamefont
  {Yin}}, \bibinfo {author} {\bibfnamefont {N.}~\bibnamefont {Zhao}},\ and\
  \bibinfo {author} {\bibfnamefont {T.}~\bibnamefont {Li}},\ }\bibfield
  {title} {\bibinfo {title} {Hybrid opto-mechanical systems with
  nitrogen-vacancy centers},\ }\href@noop {} {\bibfield  {journal} {\bibinfo
  {journal} {Science China Physics, Mechanics \& Astronomy}\ }\textbf {\bibinfo
  {volume} {58}},\ \bibinfo {pages} {1} (\bibinfo {year} {2015})}\BibitemShut
  {NoStop}%
\bibitem [{\citenamefont {Sukachev}\ \emph {et~al.}(2017)\citenamefont
  {Sukachev}, \citenamefont {Sipahigil}, \citenamefont {Nguyen}, \citenamefont
  {Bhaskar}, \citenamefont {Evans}, \citenamefont {Jelezko},\ and\
  \citenamefont {Lukin}}]{PhysRevLett.119.223602}%
  \BibitemOpen
  \bibfield  {author} {\bibinfo {author} {\bibfnamefont {D.~D.}\ \bibnamefont
  {Sukachev}}, \bibinfo {author} {\bibfnamefont {A.}~\bibnamefont {Sipahigil}},
  \bibinfo {author} {\bibfnamefont {C.~T.}\ \bibnamefont {Nguyen}}, \bibinfo
  {author} {\bibfnamefont {M.~K.}\ \bibnamefont {Bhaskar}}, \bibinfo {author}
  {\bibfnamefont {R.~E.}\ \bibnamefont {Evans}}, \bibinfo {author}
  {\bibfnamefont {F.}~\bibnamefont {Jelezko}},\ and\ \bibinfo {author}
  {\bibfnamefont {M.~D.}\ \bibnamefont {Lukin}},\ }\bibfield  {title} {\bibinfo
  {title} {Silicon-vacancy spin qubit in diamond: A quantum memory exceeding 10
  ms with single-shot state readout},\ }\href
  {https://doi.org/10.1103/PhysRevLett.119.223602} {\bibfield  {journal}
  {\bibinfo  {journal} {Phys. Rev. Lett.}\ }\textbf {\bibinfo {volume} {119}},\
  \bibinfo {pages} {223602} (\bibinfo {year} {2017})}\BibitemShut {NoStop}%
\bibitem [{\citenamefont {{Becker}}\ \emph {et~al.}(2016)\citenamefont
  {{Becker}}, \citenamefont {{G{\"o}rlitz}}, \citenamefont {{Arend}},
  \citenamefont {{Markham}},\ and\ \citenamefont
  {{Becher}}}]{2016NatCo_Becker}%
  \BibitemOpen
  \bibfield  {author} {\bibinfo {author} {\bibfnamefont {J.~N.}\ \bibnamefont
  {{Becker}}}, \bibinfo {author} {\bibfnamefont {J.}~\bibnamefont
  {{G{\"o}rlitz}}}, \bibinfo {author} {\bibfnamefont {C.}~\bibnamefont
  {{Arend}}}, \bibinfo {author} {\bibfnamefont {M.}~\bibnamefont {{Markham}}},\
  and\ \bibinfo {author} {\bibfnamefont {C.}~\bibnamefont {{Becher}}},\
  }\bibfield  {title} {\bibinfo {title} {{Ultrafast all-optical coherent
  control of single silicon vacancy colour centres in diamond}},\ }\href
  {https://doi.org/10.1038/ncomms13512} {\bibfield  {journal} {\bibinfo
  {journal} {Nat. Commun.}\ }\textbf {\bibinfo {volume} {7}},\ \bibinfo {eid}
  {13512} (\bibinfo {year} {2016})}\BibitemShut {NoStop}%
\bibitem [{\citenamefont {Rogers}\ \emph {et~al.}(2014)\citenamefont {Rogers},
  \citenamefont {Jahnke}, \citenamefont {Metsch}, \citenamefont {Sipahigil},
  \citenamefont {Binder}, \citenamefont {Teraji}, \citenamefont {Sumiya},
  \citenamefont {Isoya}, \citenamefont {Lukin}, \citenamefont {Hemmer},\ and\
  \citenamefont {Jelezko}}]{PhysRevLett.113.263602}%
  \BibitemOpen
  \bibfield  {author} {\bibinfo {author} {\bibfnamefont {L.~J.}\ \bibnamefont
  {Rogers}}, \bibinfo {author} {\bibfnamefont {K.~D.}\ \bibnamefont {Jahnke}},
  \bibinfo {author} {\bibfnamefont {M.~H.}\ \bibnamefont {Metsch}}, \bibinfo
  {author} {\bibfnamefont {A.}~\bibnamefont {Sipahigil}}, \bibinfo {author}
  {\bibfnamefont {J.~M.}\ \bibnamefont {Binder}}, \bibinfo {author}
  {\bibfnamefont {T.}~\bibnamefont {Teraji}}, \bibinfo {author} {\bibfnamefont
  {H.}~\bibnamefont {Sumiya}}, \bibinfo {author} {\bibfnamefont
  {J.}~\bibnamefont {Isoya}}, \bibinfo {author} {\bibfnamefont {M.~D.}\
  \bibnamefont {Lukin}}, \bibinfo {author} {\bibfnamefont {P.}~\bibnamefont
  {Hemmer}},\ and\ \bibinfo {author} {\bibfnamefont {F.}~\bibnamefont
  {Jelezko}},\ }\bibfield  {title} {\bibinfo {title} {All-optical
  initialization, readout, and coherent preparation of single silicon-vacancy
  spins in diamond},\ }\href {https://doi.org/10.1103/PhysRevLett.113.263602}
  {\bibfield  {journal} {\bibinfo  {journal} {Phys. Rev. Lett.}\ }\textbf
  {\bibinfo {volume} {113}},\ \bibinfo {pages} {263602} (\bibinfo {year}
  {2014})}\BibitemShut {NoStop}%
\bibitem [{\citenamefont {Becker}\ \emph {et~al.}(2018)\citenamefont {Becker},
  \citenamefont {Pingault}, \citenamefont {Gro\ss{}}, \citenamefont
  {G\"undo\ifmmode~\breve{g}\else \u{g}\fi{}an}, \citenamefont {Kukharchyk},
  \citenamefont {Markham}, \citenamefont {Edmonds}, \citenamefont {Atat\"ure},
  \citenamefont {Bushev},\ and\ \citenamefont
  {Becher}}]{PhysRevLett.120.053603}%
  \BibitemOpen
  \bibfield  {author} {\bibinfo {author} {\bibfnamefont {J.~N.}\ \bibnamefont
  {Becker}}, \bibinfo {author} {\bibfnamefont {B.}~\bibnamefont {Pingault}},
  \bibinfo {author} {\bibfnamefont {D.}~\bibnamefont {Gro\ss{}}}, \bibinfo
  {author} {\bibfnamefont {M.}~\bibnamefont {G\"undo\ifmmode~\breve{g}\else
  \u{g}\fi{}an}}, \bibinfo {author} {\bibfnamefont {N.}~\bibnamefont
  {Kukharchyk}}, \bibinfo {author} {\bibfnamefont {M.}~\bibnamefont {Markham}},
  \bibinfo {author} {\bibfnamefont {A.}~\bibnamefont {Edmonds}}, \bibinfo
  {author} {\bibfnamefont {M.}~\bibnamefont {Atat\"ure}}, \bibinfo {author}
  {\bibfnamefont {P.}~\bibnamefont {Bushev}},\ and\ \bibinfo {author}
  {\bibfnamefont {C.}~\bibnamefont {Becher}},\ }\bibfield  {title} {\bibinfo
  {title} {All-optical control of the silicon-vacancy spin in diamond at
  millikelvin temperatures},\ }\href
  {https://doi.org/10.1103/PhysRevLett.120.053603} {\bibfield  {journal}
  {\bibinfo  {journal} {Phys. Rev. Lett.}\ }\textbf {\bibinfo {volume} {120}},\
  \bibinfo {pages} {053603} (\bibinfo {year} {2018})}\BibitemShut {NoStop}%
\bibitem [{\citenamefont {{Pingault}}\ \emph {et~al.}(2017)\citenamefont
  {{Pingault}}, \citenamefont {{Jarausch}}, \citenamefont {{Hepp}},
  \citenamefont {{Klintberg}}, \citenamefont {{Becker}}, \citenamefont
  {{Markham}}, \citenamefont {{Becher}},\ and\ \citenamefont
  {{Atat{\"u}re}}}]{2017NatCo_Pingault}%
  \BibitemOpen
  \bibfield  {author} {\bibinfo {author} {\bibfnamefont {B.}~\bibnamefont
  {{Pingault}}}, \bibinfo {author} {\bibfnamefont {D.-D.}\ \bibnamefont
  {{Jarausch}}}, \bibinfo {author} {\bibfnamefont {C.}~\bibnamefont {{Hepp}}},
  \bibinfo {author} {\bibfnamefont {L.}~\bibnamefont {{Klintberg}}}, \bibinfo
  {author} {\bibfnamefont {J.~N.}\ \bibnamefont {{Becker}}}, \bibinfo {author}
  {\bibfnamefont {M.}~\bibnamefont {{Markham}}}, \bibinfo {author}
  {\bibfnamefont {C.}~\bibnamefont {{Becher}}},\ and\ \bibinfo {author}
  {\bibfnamefont {M.}~\bibnamefont {{Atat{\"u}re}}},\ }\bibfield  {title}
  {\bibinfo {title} {{Coherent control of the silicon-vacancy spin in
  diamond}},\ }\href {https://doi.org/10.1038/ncomms15579} {\bibfield
  {journal} {\bibinfo  {journal} {Nat. Commun.}\ }\textbf {\bibinfo {volume}
  {8}},\ \bibinfo {eid} {15579} (\bibinfo {year} {2017})}\BibitemShut {NoStop}%
\bibitem [{\citenamefont {Jahnke}\ \emph {et~al.}(2015)\citenamefont {Jahnke},
  \citenamefont {Sipahigil}, \citenamefont {Binder}, \citenamefont {Doherty},
  \citenamefont {Metsch}, \citenamefont {Rogers}, \citenamefont {Manson},
  \citenamefont {Lukin},\ and\ \citenamefont {Jelezko}}]{Jahnke_2015}%
  \BibitemOpen
  \bibfield  {author} {\bibinfo {author} {\bibfnamefont {K.~D.}\ \bibnamefont
  {Jahnke}}, \bibinfo {author} {\bibfnamefont {A.}~\bibnamefont {Sipahigil}},
  \bibinfo {author} {\bibfnamefont {J.~M.}\ \bibnamefont {Binder}}, \bibinfo
  {author} {\bibfnamefont {M.~W.}\ \bibnamefont {Doherty}}, \bibinfo {author}
  {\bibfnamefont {M.}~\bibnamefont {Metsch}}, \bibinfo {author} {\bibfnamefont
  {L.~J.}\ \bibnamefont {Rogers}}, \bibinfo {author} {\bibfnamefont {N.~B.}\
  \bibnamefont {Manson}}, \bibinfo {author} {\bibfnamefont {M.~D.}\
  \bibnamefont {Lukin}},\ and\ \bibinfo {author} {\bibfnamefont
  {F.}~\bibnamefont {Jelezko}},\ }\bibfield  {title} {\bibinfo {title}
  {Electron{\textendash}phonon processes of the silicon-vacancy centre in
  diamond},\ }\href {https://doi.org/10.1088/1367-2630/17/4/043011} {\bibfield
  {journal} {\bibinfo  {journal} {New J. Phys.}\ }\textbf {\bibinfo {volume}
  {17}},\ \bibinfo {pages} {043011} (\bibinfo {year} {2015})}\BibitemShut
  {NoStop}%
\bibitem [{\citenamefont {{Sohn}}\ \emph {et~al.}(2018)\citenamefont {{Sohn}},
  \citenamefont {{Meesala}}, \citenamefont {{Pingault}}, \citenamefont
  {{Atikian}}, \citenamefont {{Holzgrafe}}, \citenamefont {{G{\"u}ndo{\v
  g}an}}, \citenamefont {{Stavrakas}}, \citenamefont {{Stanley}}, \citenamefont
  {{Sipahigil}}, \citenamefont {{Choi}}, \citenamefont {{Zhang}}, \citenamefont
  {{Pacheco}}, \citenamefont {{Abraham}}, \citenamefont {{Bielejec}},
  \citenamefont {{Lukin}}, \citenamefont {{Atat{\"u}re}},\ and\ \citenamefont
  {{Lon{\v c}ar}}}]{Nature.Comm-SiV-strain}%
  \BibitemOpen
  \bibfield  {author} {\bibinfo {author} {\bibfnamefont {Y.-I.}\ \bibnamefont
  {{Sohn}}}, \bibinfo {author} {\bibfnamefont {S.}~\bibnamefont {{Meesala}}},
  \bibinfo {author} {\bibfnamefont {B.}~\bibnamefont {{Pingault}}}, \bibinfo
  {author} {\bibfnamefont {H.~A.}\ \bibnamefont {{Atikian}}}, \bibinfo {author}
  {\bibfnamefont {J.}~\bibnamefont {{Holzgrafe}}}, \bibinfo {author}
  {\bibfnamefont {M.}~\bibnamefont {{G{\"u}ndo{\v g}an}}}, \bibinfo {author}
  {\bibfnamefont {C.}~\bibnamefont {{Stavrakas}}}, \bibinfo {author}
  {\bibfnamefont {M.~J.}\ \bibnamefont {{Stanley}}}, \bibinfo {author}
  {\bibfnamefont {A.}~\bibnamefont {{Sipahigil}}}, \bibinfo {author}
  {\bibfnamefont {J.}~\bibnamefont {{Choi}}}, \bibinfo {author} {\bibfnamefont
  {M.}~\bibnamefont {{Zhang}}}, \bibinfo {author} {\bibfnamefont {J.~L.}\
  \bibnamefont {{Pacheco}}}, \bibinfo {author} {\bibfnamefont {J.}~\bibnamefont
  {{Abraham}}}, \bibinfo {author} {\bibfnamefont {E.}~\bibnamefont
  {{Bielejec}}}, \bibinfo {author} {\bibfnamefont {M.~D.}\ \bibnamefont
  {{Lukin}}}, \bibinfo {author} {\bibfnamefont {M.}~\bibnamefont
  {{Atat{\"u}re}}},\ and\ \bibinfo {author} {\bibfnamefont {M.}~\bibnamefont
  {{Lon{\v c}ar}}},\ }\bibfield  {title} {\bibinfo {title} {{Controlling the
  coherence of a diamond spin qubit through its strain environment}},\ }\href
  {https://doi.org/10.1038/s41467-018-04340-3} {\bibfield  {journal} {\bibinfo
  {journal} {Nat. Commun.}\ }\textbf {\bibinfo {volume} {9}},\ \bibinfo {eid}
  {2012} (\bibinfo {year} {2018})}\BibitemShut {NoStop}%
\bibitem [{\citenamefont {Kepesidis}\ \emph {et~al.}(2016)\citenamefont
  {Kepesidis}, \citenamefont {Lemonde}, \citenamefont {Norambuena},
  \citenamefont {Maze},\ and\ \citenamefont {Rabl}}]{PhysRevB.94.214115}%
  \BibitemOpen
  \bibfield  {author} {\bibinfo {author} {\bibfnamefont {K.~V.}\ \bibnamefont
  {Kepesidis}}, \bibinfo {author} {\bibfnamefont {M.-A.}\ \bibnamefont
  {Lemonde}}, \bibinfo {author} {\bibfnamefont {A.}~\bibnamefont {Norambuena}},
  \bibinfo {author} {\bibfnamefont {J.~R.}\ \bibnamefont {Maze}},\ and\
  \bibinfo {author} {\bibfnamefont {P.}~\bibnamefont {Rabl}},\ }\bibfield
  {title} {\bibinfo {title} {Cooling phonons with phonons: Acoustic reservoir
  engineering with silicon-vacancy centers in diamond},\ }\href
  {https://doi.org/10.1103/PhysRevB.94.214115} {\bibfield  {journal} {\bibinfo
  {journal} {Phys. Rev. B}\ }\textbf {\bibinfo {volume} {94}},\ \bibinfo
  {pages} {214115} (\bibinfo {year} {2016})}\BibitemShut {NoStop}%
\bibitem [{\citenamefont {Meesala}\ \emph {et~al.}(2018)\citenamefont
  {Meesala}, \citenamefont {Sohn}, \citenamefont {Pingault}, \citenamefont
  {Shao}, \citenamefont {Atikian}, \citenamefont {Holzgrafe}, \citenamefont
  {G\"undo\ifmmode~\breve{g}\else \u{g}\fi{}an}, \citenamefont {Stavrakas},
  \citenamefont {Sipahigil}, \citenamefont {Chia}, \citenamefont {Evans},
  \citenamefont {Burek}, \citenamefont {Zhang}, \citenamefont {Wu},
  \citenamefont {Pacheco}, \citenamefont {Abraham}, \citenamefont {Bielejec},
  \citenamefont {Lukin}, \citenamefont {Atat\"ure},\ and\ \citenamefont
  {Lon\ifmmode~\check{c}\else \v{c}\fi{}ar}}]{PhysRevB.97.205444}%
  \BibitemOpen
  \bibfield  {author} {\bibinfo {author} {\bibfnamefont {S.}~\bibnamefont
  {Meesala}}, \bibinfo {author} {\bibfnamefont {Y.-I.}\ \bibnamefont {Sohn}},
  \bibinfo {author} {\bibfnamefont {B.}~\bibnamefont {Pingault}}, \bibinfo
  {author} {\bibfnamefont {L.}~\bibnamefont {Shao}}, \bibinfo {author}
  {\bibfnamefont {H.~A.}\ \bibnamefont {Atikian}}, \bibinfo {author}
  {\bibfnamefont {J.}~\bibnamefont {Holzgrafe}}, \bibinfo {author}
  {\bibfnamefont {M.}~\bibnamefont {G\"undo\ifmmode~\breve{g}\else
  \u{g}\fi{}an}}, \bibinfo {author} {\bibfnamefont {C.}~\bibnamefont
  {Stavrakas}}, \bibinfo {author} {\bibfnamefont {A.}~\bibnamefont
  {Sipahigil}}, \bibinfo {author} {\bibfnamefont {C.}~\bibnamefont {Chia}},
  \bibinfo {author} {\bibfnamefont {R.}~\bibnamefont {Evans}}, \bibinfo
  {author} {\bibfnamefont {M.~J.}\ \bibnamefont {Burek}}, \bibinfo {author}
  {\bibfnamefont {M.}~\bibnamefont {Zhang}}, \bibinfo {author} {\bibfnamefont
  {L.}~\bibnamefont {Wu}}, \bibinfo {author} {\bibfnamefont {J.~L.}\
  \bibnamefont {Pacheco}}, \bibinfo {author} {\bibfnamefont {J.}~\bibnamefont
  {Abraham}}, \bibinfo {author} {\bibfnamefont {E.}~\bibnamefont {Bielejec}},
  \bibinfo {author} {\bibfnamefont {M.~D.}\ \bibnamefont {Lukin}}, \bibinfo
  {author} {\bibfnamefont {M.}~\bibnamefont {Atat\"ure}},\ and\ \bibinfo
  {author} {\bibfnamefont {M.}~\bibnamefont {Lon\ifmmode~\check{c}\else
  \v{c}\fi{}ar}},\ }\bibfield  {title} {\bibinfo {title} {Strain engineering of
  the silicon-vacancy center in diamond},\ }\href
  {https://doi.org/10.1103/PhysRevB.97.205444} {\bibfield  {journal} {\bibinfo
  {journal} {Phys. Rev. B}\ }\textbf {\bibinfo {volume} {97}},\ \bibinfo
  {pages} {205444} (\bibinfo {year} {2018})}\BibitemShut {NoStop}%
\bibitem [{\citenamefont {Li}\ \emph {et~al.}()\citenamefont {Li},
  \citenamefont {Li},\ and\ \citenamefont {Nori}}]{arxiv-1901-04650}%
  \BibitemOpen
  \bibfield  {author} {\bibinfo {author} {\bibfnamefont {P.-B.}\ \bibnamefont
  {Li}}, \bibinfo {author} {\bibfnamefont {X.-X.}\ \bibnamefont {Li}},\ and\
  \bibinfo {author} {\bibfnamefont {F.}~\bibnamefont {Nori}},\ }\bibfield
  {title} {\bibinfo {title} {Band-gap-engineered spin-phonon, and spin-spin
  interactions with defect centers in diamond coupled to phononic crystals},\
  }\href {https://doi.org/arxiv.org/abs/1901.04650} {\ }\Eprint
  {https://arxiv.org/abs/1901.04650} {arXiv:1901.04650} \BibitemShut {NoStop}%
\bibitem [{\citenamefont {Lemonde}\ \emph {et~al.}(2018)\citenamefont
  {Lemonde}, \citenamefont {Meesala}, \citenamefont {Sipahigil}, \citenamefont
  {Schuetz}, \citenamefont {Lukin}, \citenamefont {Loncar},\ and\ \citenamefont
  {Rabl}}]{PhysRevLett.120.213603}%
  \BibitemOpen
  \bibfield  {author} {\bibinfo {author} {\bibfnamefont {M.-A.}\ \bibnamefont
  {Lemonde}}, \bibinfo {author} {\bibfnamefont {S.}~\bibnamefont {Meesala}},
  \bibinfo {author} {\bibfnamefont {A.}~\bibnamefont {Sipahigil}}, \bibinfo
  {author} {\bibfnamefont {M.~J.~A.}\ \bibnamefont {Schuetz}}, \bibinfo
  {author} {\bibfnamefont {M.~D.}\ \bibnamefont {Lukin}}, \bibinfo {author}
  {\bibfnamefont {M.}~\bibnamefont {Loncar}},\ and\ \bibinfo {author}
  {\bibfnamefont {P.}~\bibnamefont {Rabl}},\ }\bibfield  {title} {\bibinfo
  {title} {Phonon networks with silicon-vacancy centers in diamond
  waveguides},\ }\href {https://doi.org/10.1103/PhysRevLett.120.213603}
  {\bibfield  {journal} {\bibinfo  {journal} {Phys. Rev. Lett.}\ }\textbf
  {\bibinfo {volume} {120}},\ \bibinfo {pages} {213603} (\bibinfo {year}
  {2018})}\BibitemShut {NoStop}%
\bibitem [{\citenamefont {Lee}\ \emph {et~al.}(2017)\citenamefont {Lee},
  \citenamefont {Lee}, \citenamefont {Cady}, \citenamefont {Ovartchaiyapong},\
  and\ \citenamefont {Jayich}}]{Lee_2017}%
  \BibitemOpen
  \bibfield  {author} {\bibinfo {author} {\bibfnamefont {D.}~\bibnamefont
  {Lee}}, \bibinfo {author} {\bibfnamefont {K.~W.}\ \bibnamefont {Lee}},
  \bibinfo {author} {\bibfnamefont {J.~V.}\ \bibnamefont {Cady}}, \bibinfo
  {author} {\bibfnamefont {P.}~\bibnamefont {Ovartchaiyapong}},\ and\ \bibinfo
  {author} {\bibfnamefont {A.~C.~B.}\ \bibnamefont {Jayich}},\ }\bibfield
  {title} {\bibinfo {title} {Topical review: spins and mechanics in diamond},\
  }\href {https://doi.org/10.1088/2040-8986/aa52cd} {\bibfield  {journal}
  {\bibinfo  {journal} {J. Opt.}\ }\textbf {\bibinfo {volume} {19}},\ \bibinfo
  {pages} {033001} (\bibinfo {year} {2017})}\BibitemShut {NoStop}%
\bibitem [{\citenamefont {{Barfuss}}\ \emph {et~al.}(2015)\citenamefont
  {{Barfuss}}, \citenamefont {{Teissier}}, \citenamefont {{Neu}}, \citenamefont
  {{Nunnenkamp}},\ and\ \citenamefont {{Maletinsky}}}]{NatPhy_NV_AC}%
  \BibitemOpen
  \bibfield  {author} {\bibinfo {author} {\bibfnamefont {A.}~\bibnamefont
  {{Barfuss}}}, \bibinfo {author} {\bibfnamefont {J.}~\bibnamefont
  {{Teissier}}}, \bibinfo {author} {\bibfnamefont {E.}~\bibnamefont {{Neu}}},
  \bibinfo {author} {\bibfnamefont {A.}~\bibnamefont {{Nunnenkamp}}},\ and\
  \bibinfo {author} {\bibfnamefont {P.}~\bibnamefont {{Maletinsky}}},\
  }\bibfield  {title} {\bibinfo {title} {{Strong mechanical driving of a single
  electron spin}},\ }\href {https://doi.org/10.1038/nphys3411} {\bibfield
  {journal} {\bibinfo  {journal} {Nature Phys.}\ }\textbf {\bibinfo {volume}
  {11}},\ \bibinfo {pages} {820} (\bibinfo {year} {2015})}\BibitemShut
  {NoStop}%
\bibitem [{\citenamefont {Teissier}\ \emph {et~al.}(2014)\citenamefont
  {Teissier}, \citenamefont {Barfuss}, \citenamefont {Appel}, \citenamefont
  {Neu},\ and\ \citenamefont {Maletinsky}}]{PhysRevLett.113.020503}%
  \BibitemOpen
  \bibfield  {author} {\bibinfo {author} {\bibfnamefont {J.}~\bibnamefont
  {Teissier}}, \bibinfo {author} {\bibfnamefont {A.}~\bibnamefont {Barfuss}},
  \bibinfo {author} {\bibfnamefont {P.}~\bibnamefont {Appel}}, \bibinfo
  {author} {\bibfnamefont {E.}~\bibnamefont {Neu}},\ and\ \bibinfo {author}
  {\bibfnamefont {P.}~\bibnamefont {Maletinsky}},\ }\bibfield  {title}
  {\bibinfo {title} {Strain coupling of a nitrogen-vacancy center spin to a
  diamond mechanical oscillator},\ }\href
  {https://doi.org/10.1103/PhysRevLett.113.020503} {\bibfield  {journal}
  {\bibinfo  {journal} {Phys. Rev. Lett.}\ }\textbf {\bibinfo {volume} {113}},\
  \bibinfo {pages} {020503} (\bibinfo {year} {2014})}\BibitemShut {NoStop}%
\bibitem [{\citenamefont {{Whiteley}}\ \emph {et~al.}(2019)\citenamefont
  {{Whiteley}}, \citenamefont {{Wolfowicz}}, \citenamefont {{Anderson}},
  \citenamefont {{Bourassa}}, \citenamefont {{Ma}}, \citenamefont {{Ye}},
  \citenamefont {{Koolstra}}, \citenamefont {{Satzinger}}, \citenamefont
  {{Holt}}, \citenamefont {{Heremans}}, \citenamefont {{Cleland}},
  \citenamefont {{Schuster}}, \citenamefont {{Galli}},\ and\ \citenamefont
  {{Awschalom}}}]{2018arXiv180410996W}%
  \BibitemOpen
  \bibfield  {author} {\bibinfo {author} {\bibfnamefont {S.~J.}\ \bibnamefont
  {{Whiteley}}}, \bibinfo {author} {\bibfnamefont {G.}~\bibnamefont
  {{Wolfowicz}}}, \bibinfo {author} {\bibfnamefont {C.~P.}\ \bibnamefont
  {{Anderson}}}, \bibinfo {author} {\bibfnamefont {A.}~\bibnamefont
  {{Bourassa}}}, \bibinfo {author} {\bibfnamefont {H.}~\bibnamefont {{Ma}}},
  \bibinfo {author} {\bibfnamefont {M.}~\bibnamefont {{Ye}}}, \bibinfo {author}
  {\bibfnamefont {G.}~\bibnamefont {{Koolstra}}}, \bibinfo {author}
  {\bibfnamefont {K.~J.}\ \bibnamefont {{Satzinger}}}, \bibinfo {author}
  {\bibfnamefont {M.~V.}\ \bibnamefont {{Holt}}}, \bibinfo {author}
  {\bibfnamefont {F.~J.}\ \bibnamefont {{Heremans}}}, \bibinfo {author}
  {\bibfnamefont {A.~N.}\ \bibnamefont {{Cleland}}}, \bibinfo {author}
  {\bibfnamefont {D.~I.}\ \bibnamefont {{Schuster}}}, \bibinfo {author}
  {\bibfnamefont {G.}~\bibnamefont {{Galli}}},\ and\ \bibinfo {author}
  {\bibfnamefont {D.~D.}\ \bibnamefont {{Awschalom}}},\ }\bibfield  {title}
  {\bibinfo {title} {{Spin-phonon interactions in silicon carbide addressed by
  Gaussian acoustics}},\ }\href
  {https://journals.aps.org/prl/pdf/10.1103/PhysRevLett.120.213603} {\bibfield
  {journal} {\bibinfo  {journal} {Nat. Phys.}\ }\textbf {\bibinfo {volume}
  {15}},\ \bibinfo {pages} {490} (\bibinfo {year} {2019})}\BibitemShut
  {NoStop}%
\bibitem [{\citenamefont {{Barson}}\ \emph {et~al.}(2017)\citenamefont
  {{Barson}}, \citenamefont {{Peddibhotla}}, \citenamefont {{Ovartchaiyapong}},
  \citenamefont {{Ganesan}}, \citenamefont {{Taylor}}, \citenamefont
  {{Gebert}}, \citenamefont {{Mielens}}, \citenamefont {{Koslowski}},
  \citenamefont {{Simpson}}, \citenamefont {{McGuinness}}, \citenamefont
  {{McCallum}}, \citenamefont {{Prawer}}, \citenamefont {{Onoda}},
  \citenamefont {{Ohshima}}, \citenamefont {{Bleszynski Jayich}}, \citenamefont
  {{Jelezko}}, \citenamefont {{Manson}},\ and\ \citenamefont
  {{Doherty}}}]{NL-17-NV}%
  \BibitemOpen
  \bibfield  {author} {\bibinfo {author} {\bibfnamefont {M.~S.~J.}\
  \bibnamefont {{Barson}}}, \bibinfo {author} {\bibfnamefont {P.}~\bibnamefont
  {{Peddibhotla}}}, \bibinfo {author} {\bibfnamefont {P.}~\bibnamefont
  {{Ovartchaiyapong}}}, \bibinfo {author} {\bibfnamefont {K.}~\bibnamefont
  {{Ganesan}}}, \bibinfo {author} {\bibfnamefont {R.~L.}\ \bibnamefont
  {{Taylor}}}, \bibinfo {author} {\bibfnamefont {M.}~\bibnamefont {{Gebert}}},
  \bibinfo {author} {\bibfnamefont {Z.}~\bibnamefont {{Mielens}}}, \bibinfo
  {author} {\bibfnamefont {B.}~\bibnamefont {{Koslowski}}}, \bibinfo {author}
  {\bibfnamefont {D.~A.}\ \bibnamefont {{Simpson}}}, \bibinfo {author}
  {\bibfnamefont {L.~P.}\ \bibnamefont {{McGuinness}}}, \bibinfo {author}
  {\bibfnamefont {J.}~\bibnamefont {{McCallum}}}, \bibinfo {author}
  {\bibfnamefont {S.}~\bibnamefont {{Prawer}}}, \bibinfo {author}
  {\bibfnamefont {S.}~\bibnamefont {{Onoda}}}, \bibinfo {author} {\bibfnamefont
  {T.}~\bibnamefont {{Ohshima}}}, \bibinfo {author} {\bibfnamefont {A.~C.}\
  \bibnamefont {{Bleszynski Jayich}}}, \bibinfo {author} {\bibfnamefont
  {F.}~\bibnamefont {{Jelezko}}}, \bibinfo {author} {\bibfnamefont {N.~B.}\
  \bibnamefont {{Manson}}},\ and\ \bibinfo {author} {\bibfnamefont {M.~W.}\
  \bibnamefont {{Doherty}}},\ }\bibfield  {title} {\bibinfo {title}
  {{Nanomechanical Sensing Using Spins in Diamond}},\ }\href
  {https://doi.org/10.1021/acs.nanolett.6b04544} {\bibfield  {journal}
  {\bibinfo  {journal} {Nano Lett.}\ }\textbf {\bibinfo {volume} {17}},\
  \bibinfo {pages} {1496} (\bibinfo {year} {2017})}\BibitemShut {NoStop}%
\bibitem [{\citenamefont {{Falk}}\ \emph {et~al.}(2013)\citenamefont {{Falk}},
  \citenamefont {{Buckley}}, \citenamefont {{Calusine}}, \citenamefont
  {{Koehl}}, \citenamefont {{Dobrovitski}}, \citenamefont {{Politi}},
  \citenamefont {{Zorman}}, \citenamefont {{Feng}},\ and\ \citenamefont
  {{Awschalom}}}]{2013NatC-4E1819F}%
  \BibitemOpen
  \bibfield  {author} {\bibinfo {author} {\bibfnamefont {A.~L.}\ \bibnamefont
  {{Falk}}}, \bibinfo {author} {\bibfnamefont {B.~B.}\ \bibnamefont
  {{Buckley}}}, \bibinfo {author} {\bibfnamefont {G.}~\bibnamefont
  {{Calusine}}}, \bibinfo {author} {\bibfnamefont {W.~F.}\ \bibnamefont
  {{Koehl}}}, \bibinfo {author} {\bibfnamefont {V.~V.}\ \bibnamefont
  {{Dobrovitski}}}, \bibinfo {author} {\bibfnamefont {A.}~\bibnamefont
  {{Politi}}}, \bibinfo {author} {\bibfnamefont {C.~A.}\ \bibnamefont
  {{Zorman}}}, \bibinfo {author} {\bibfnamefont {P.~X.-L.}\ \bibnamefont
  {{Feng}}},\ and\ \bibinfo {author} {\bibfnamefont {D.~D.}\ \bibnamefont
  {{Awschalom}}},\ }\bibfield  {title} {\bibinfo {title} {{Polytype control of
  spin qubits in silicon carbide}},\ }\href
  {https://doi.org/10.1038/ncomms2854} {\bibfield  {journal} {\bibinfo
  {journal} {Nat. Commun.}\ }\textbf {\bibinfo {volume} {4}},\ \bibinfo {eid}
  {1819} (\bibinfo {year} {2013})}\BibitemShut {NoStop}%
\bibitem [{\citenamefont {Kitagawa}\ and\ \citenamefont
  {Ueda}(1993)}]{PhysRevA.47.5138}%
  \BibitemOpen
  \bibfield  {author} {\bibinfo {author} {\bibfnamefont {M.}~\bibnamefont
  {Kitagawa}}\ and\ \bibinfo {author} {\bibfnamefont {M.}~\bibnamefont
  {Ueda}},\ }\bibfield  {title} {\bibinfo {title} {Squeezed spin states},\
  }\href {https://doi.org/10.1103/PhysRevA.47.5138} {\bibfield  {journal}
  {\bibinfo  {journal} {Phys. Rev. A}\ }\textbf {\bibinfo {volume} {47}},\
  \bibinfo {pages} {5138} (\bibinfo {year} {1993})}\BibitemShut {NoStop}%
\bibitem [{\citenamefont {Ma}\ \emph {et~al.}(2011)\citenamefont {Ma},
  \citenamefont {Wang}, \citenamefont {Sun},\ and\ \citenamefont
  {Nori}}]{MA201189}%
  \BibitemOpen
  \bibfield  {author} {\bibinfo {author} {\bibfnamefont {J.}~\bibnamefont
  {Ma}}, \bibinfo {author} {\bibfnamefont {X.}~\bibnamefont {Wang}}, \bibinfo
  {author} {\bibfnamefont {C.~P.}\ \bibnamefont {Sun}},\ and\ \bibinfo {author}
  {\bibfnamefont {F.}~\bibnamefont {Nori}},\ }\bibfield  {title} {\bibinfo
  {title} {Quantum spin squeezing},\ }\href
  {https://doi.org/https://doi.org/10.1016/j.physrep.2011.08.003} {\bibfield
  {journal} {\bibinfo  {journal} {Phys. Rep.}\ }\textbf {\bibinfo {volume}
  {509}},\ \bibinfo {pages} {89} (\bibinfo {year} {2011})}\BibitemShut
  {NoStop}%
\bibitem [{\citenamefont {Jin}\ \emph {et~al.}(2009)\citenamefont {Jin},
  \citenamefont {Liu},\ and\ \citenamefont {Liu}}]{Jin_2009}%
  \BibitemOpen
  \bibfield  {author} {\bibinfo {author} {\bibfnamefont {G.-R.}\ \bibnamefont
  {Jin}}, \bibinfo {author} {\bibfnamefont {Y.-C.}\ \bibnamefont {Liu}},\ and\
  \bibinfo {author} {\bibfnamefont {W.-M.}\ \bibnamefont {Liu}},\ }\bibfield
  {title} {\bibinfo {title} {Spin squeezing in a generalized one-axis twisting
  model},\ }\href {https://doi.org/10.1088/1367-2630/11/7/073049} {\bibfield
  {journal} {\bibinfo  {journal} {New J. Phys.}\ }\textbf {\bibinfo {volume}
  {11}},\ \bibinfo {pages} {073049} (\bibinfo {year} {2009})}\BibitemShut
  {NoStop}%
\bibitem [{\citenamefont {Degen}\ \emph {et~al.}(2017)\citenamefont {Degen},
  \citenamefont {Reinhard},\ and\ \citenamefont
  {Cappellaro}}]{RevModPhys.89.035002}%
  \BibitemOpen
  \bibfield  {author} {\bibinfo {author} {\bibfnamefont {C.~L.}\ \bibnamefont
  {Degen}}, \bibinfo {author} {\bibfnamefont {F.}~\bibnamefont {Reinhard}},\
  and\ \bibinfo {author} {\bibfnamefont {P.}~\bibnamefont {Cappellaro}},\
  }\bibfield  {title} {\bibinfo {title} {Quantum sensing},\ }\href
  {https://doi.org/10.1103/RevModPhys.89.035002} {\bibfield  {journal}
  {\bibinfo  {journal} {Rev. Mod. Phys.}\ }\textbf {\bibinfo {volume} {89}},\
  \bibinfo {pages} {035002} (\bibinfo {year} {2017})}\BibitemShut {NoStop}%
\bibitem [{\citenamefont {{Luo}}\ \emph {et~al.}(2017)\citenamefont {{Luo}},
  \citenamefont {{Zou}}, \citenamefont {{Wu}}, \citenamefont {{Liu}},
  \citenamefont {{Han}}, \citenamefont {{Tey}},\ and\ \citenamefont
  {{You}}}]{Science-355-620}%
  \BibitemOpen
  \bibfield  {author} {\bibinfo {author} {\bibfnamefont {X.-Y.}\ \bibnamefont
  {{Luo}}}, \bibinfo {author} {\bibfnamefont {Y.-Q.}\ \bibnamefont {{Zou}}},
  \bibinfo {author} {\bibfnamefont {L.-N.}\ \bibnamefont {{Wu}}}, \bibinfo
  {author} {\bibfnamefont {Q.}~\bibnamefont {{Liu}}}, \bibinfo {author}
  {\bibfnamefont {M.-F.}\ \bibnamefont {{Han}}}, \bibinfo {author}
  {\bibfnamefont {M.~K.}\ \bibnamefont {{Tey}}},\ and\ \bibinfo {author}
  {\bibfnamefont {L.}~\bibnamefont {{You}}},\ }\bibfield  {title} {\bibinfo
  {title} {{Deterministic entanglement generation from driving through quantum
  phase transitions}},\ }\href {https://doi.org/10.1126/science.aag1106}
  {\bibfield  {journal} {\bibinfo  {journal} {Science}\ }\textbf {\bibinfo
  {volume} {355}},\ \bibinfo {pages} {620} (\bibinfo {year}
  {2017})}\BibitemShut {NoStop}%
\bibitem [{\citenamefont {{Choi}}\ \emph {et~al.}(2018)\citenamefont {{Choi}},
  \citenamefont {{Yao}},\ and\ \citenamefont {{Lukin}}}]{arXiv-1801-00042}%
  \BibitemOpen
  \bibfield  {author} {\bibinfo {author} {\bibfnamefont {S.}~\bibnamefont
  {{Choi}}}, \bibinfo {author} {\bibfnamefont {N.~Y.}\ \bibnamefont {{Yao}}},\
  and\ \bibinfo {author} {\bibfnamefont {M.~D.}\ \bibnamefont {{Lukin}}},\
  }\bibfield  {title} {\bibinfo {title} {{Quantum metrology based on strongly
  correlated matter}},\ }\href {https://arxiv.org/abs/1801.00042} {\  (\bibinfo
  {year} {2018})},\ \Eprint {https://arxiv.org/abs/1801.00042}
  {arXiv:1801.00042} \BibitemShut {NoStop}%
\bibitem [{\citenamefont {Hammerer}\ \emph {et~al.}(2010)\citenamefont
  {Hammerer}, \citenamefont {S\o{}rensen},\ and\ \citenamefont
  {Polzik}}]{RevModPhys.82.1041}%
  \BibitemOpen
  \bibfield  {author} {\bibinfo {author} {\bibfnamefont {K.}~\bibnamefont
  {Hammerer}}, \bibinfo {author} {\bibfnamefont {A.~S.}\ \bibnamefont
  {S\o{}rensen}},\ and\ \bibinfo {author} {\bibfnamefont {E.~S.}\ \bibnamefont
  {Polzik}},\ }\bibfield  {title} {\bibinfo {title} {Quantum interface between
  light and atomic ensembles},\ }\href
  {https://doi.org/10.1103/RevModPhys.82.1041} {\bibfield  {journal} {\bibinfo
  {journal} {Rev. Mod. Phys.}\ }\textbf {\bibinfo {volume} {82}},\ \bibinfo
  {pages} {1041} (\bibinfo {year} {2010})}\BibitemShut {NoStop}%
\bibitem [{\citenamefont {{Julsgaard}}\ \emph {et~al.}(2001)\citenamefont
  {{Julsgaard}}, \citenamefont {{Kozhekin}},\ and\ \citenamefont
  {{Polzik}}}]{2001Natur.413.400J}%
  \BibitemOpen
  \bibfield  {author} {\bibinfo {author} {\bibfnamefont {B.}~\bibnamefont
  {{Julsgaard}}}, \bibinfo {author} {\bibfnamefont {A.}~\bibnamefont
  {{Kozhekin}}},\ and\ \bibinfo {author} {\bibfnamefont {E.~S.}\ \bibnamefont
  {{Polzik}}},\ }\bibfield  {title} {\bibinfo {title} {{Experimental long-lived
  entanglement of two macroscopic objects}},\ }\href
  {https://doi.org/10.1038/35096524} {\bibfield  {journal} {\bibinfo  {journal}
  {\nat}\ }\textbf {\bibinfo {volume} {413}},\ \bibinfo {pages} {400} (\bibinfo
  {year} {2001})}\BibitemShut {NoStop}%
\bibitem [{\citenamefont {Vernac}\ \emph {et~al.}(2000)\citenamefont {Vernac},
  \citenamefont {Pinard},\ and\ \citenamefont
  {Giacobino}}]{PhysRevA.62.063812}%
  \BibitemOpen
  \bibfield  {author} {\bibinfo {author} {\bibfnamefont {L.}~\bibnamefont
  {Vernac}}, \bibinfo {author} {\bibfnamefont {M.}~\bibnamefont {Pinard}},\
  and\ \bibinfo {author} {\bibfnamefont {E.}~\bibnamefont {Giacobino}},\
  }\bibfield  {title} {\bibinfo {title} {Spin squeezing in two-level systems},\
  }\href {https://doi.org/10.1103/PhysRevA.62.063812} {\bibfield  {journal}
  {\bibinfo  {journal} {Phys. Rev. A}\ }\textbf {\bibinfo {volume} {62}},\
  \bibinfo {pages} {063812} (\bibinfo {year} {2000})}\BibitemShut {NoStop}%
\bibitem [{\citenamefont {Chaudhury}\ \emph {et~al.}(2007)\citenamefont
  {Chaudhury}, \citenamefont {Merkel}, \citenamefont {Herr}, \citenamefont
  {Silberfarb}, \citenamefont {Deutsch},\ and\ \citenamefont
  {Jessen}}]{PhysRevLett.99.163002}%
  \BibitemOpen
  \bibfield  {author} {\bibinfo {author} {\bibfnamefont {S.}~\bibnamefont
  {Chaudhury}}, \bibinfo {author} {\bibfnamefont {S.}~\bibnamefont {Merkel}},
  \bibinfo {author} {\bibfnamefont {T.}~\bibnamefont {Herr}}, \bibinfo {author}
  {\bibfnamefont {A.}~\bibnamefont {Silberfarb}}, \bibinfo {author}
  {\bibfnamefont {I.~H.}\ \bibnamefont {Deutsch}},\ and\ \bibinfo {author}
  {\bibfnamefont {P.~S.}\ \bibnamefont {Jessen}},\ }\bibfield  {title}
  {\bibinfo {title} {Quantum control of the hyperfine spin of a cs atom
  ensemble},\ }\href {https://doi.org/10.1103/PhysRevLett.99.163002} {\bibfield
   {journal} {\bibinfo  {journal} {Phys. Rev. Lett.}\ }\textbf {\bibinfo
  {volume} {99}},\ \bibinfo {pages} {163002} (\bibinfo {year}
  {2007})}\BibitemShut {NoStop}%
\bibitem [{\citenamefont {Inoue}\ \emph {et~al.}(2013)\citenamefont {Inoue},
  \citenamefont {Tanaka}, \citenamefont {Namiki}, \citenamefont {Sagawa},\ and\
  \citenamefont {Takahashi}}]{PhysRevLett.110.163602}%
  \BibitemOpen
  \bibfield  {author} {\bibinfo {author} {\bibfnamefont {R.}~\bibnamefont
  {Inoue}}, \bibinfo {author} {\bibfnamefont {S.-I.-R.}\ \bibnamefont
  {Tanaka}}, \bibinfo {author} {\bibfnamefont {R.}~\bibnamefont {Namiki}},
  \bibinfo {author} {\bibfnamefont {T.}~\bibnamefont {Sagawa}},\ and\ \bibinfo
  {author} {\bibfnamefont {Y.}~\bibnamefont {Takahashi}},\ }\bibfield  {title}
  {\bibinfo {title} {Unconditional quantum-noise suppression via
  measurement-based quantum feedback},\ }\href
  {https://doi.org/10.1103/PhysRevLett.110.163602} {\bibfield  {journal}
  {\bibinfo  {journal} {Phys. Rev. Lett.}\ }\textbf {\bibinfo {volume} {110}},\
  \bibinfo {pages} {163602} (\bibinfo {year} {2013})}\BibitemShut {NoStop}%
\bibitem [{\citenamefont {Fernholz}\ \emph {et~al.}(2008)\citenamefont
  {Fernholz}, \citenamefont {Krauter}, \citenamefont {Jensen}, \citenamefont
  {Sherson}, \citenamefont {S\o{}rensen},\ and\ \citenamefont
  {Polzik}}]{PhysRevLett.101.073601}%
  \BibitemOpen
  \bibfield  {author} {\bibinfo {author} {\bibfnamefont {T.}~\bibnamefont
  {Fernholz}}, \bibinfo {author} {\bibfnamefont {H.}~\bibnamefont {Krauter}},
  \bibinfo {author} {\bibfnamefont {K.}~\bibnamefont {Jensen}}, \bibinfo
  {author} {\bibfnamefont {J.~F.}\ \bibnamefont {Sherson}}, \bibinfo {author}
  {\bibfnamefont {A.~S.}\ \bibnamefont {S\o{}rensen}},\ and\ \bibinfo {author}
  {\bibfnamefont {E.~S.}\ \bibnamefont {Polzik}},\ }\bibfield  {title}
  {\bibinfo {title} {Spin squeezing of atomic ensembles via nuclear-electronic
  spin entanglement},\ }\href {https://doi.org/10.1103/PhysRevLett.101.073601}
  {\bibfield  {journal} {\bibinfo  {journal} {Phys. Rev. Lett.}\ }\textbf
  {\bibinfo {volume} {101}},\ \bibinfo {pages} {073601} (\bibinfo {year}
  {2008})}\BibitemShut {NoStop}%
\bibitem [{\citenamefont {Shindo}\ \emph {et~al.}(2004)\citenamefont {Shindo},
  \citenamefont {Chavez}, \citenamefont {Chumakov},\ and\ \citenamefont
  {Klimov}}]{Shindo_2003}%
  \BibitemOpen
  \bibfield  {author} {\bibinfo {author} {\bibfnamefont {D.}~\bibnamefont
  {Shindo}}, \bibinfo {author} {\bibfnamefont {A.}~\bibnamefont {Chavez}},
  \bibinfo {author} {\bibfnamefont {S.~M.}\ \bibnamefont {Chumakov}},\ and\
  \bibinfo {author} {\bibfnamefont {A.~B.}\ \bibnamefont {Klimov}},\ }\bibfield
   {title} {\bibinfo {title} {Dynamical squeezing enhancement in the
  off-resonant dicke model},\ }\href
  {https://doi.org/10.1088/1464-4266/6/1/006} {\bibfield  {journal} {\bibinfo
  {journal} {J. Opt. B: Quantum Semiclassical Opt.}\ }\textbf {\bibinfo
  {volume} {6}},\ \bibinfo {pages} {34} (\bibinfo {year} {2004})}\BibitemShut
  {NoStop}%
\bibitem [{\citenamefont {Deb}\ \emph {et~al.}(2006)\citenamefont {Deb},
  \citenamefont {Abdalla}, \citenamefont {Hassan},\ and\ \citenamefont
  {Nayak}}]{PhysRevA.73.053817}%
  \BibitemOpen
  \bibfield  {author} {\bibinfo {author} {\bibfnamefont {R.~N.}\ \bibnamefont
  {Deb}}, \bibinfo {author} {\bibfnamefont {M.~S.}\ \bibnamefont {Abdalla}},
  \bibinfo {author} {\bibfnamefont {S.~S.}\ \bibnamefont {Hassan}},\ and\
  \bibinfo {author} {\bibfnamefont {N.}~\bibnamefont {Nayak}},\ }\bibfield
  {title} {\bibinfo {title} {Spin squeezing and entanglement in a dispersive
  cavity},\ }\href {https://doi.org/10.1103/PhysRevA.73.053817} {\bibfield
  {journal} {\bibinfo  {journal} {Phys. Rev. A}\ }\textbf {\bibinfo {volume}
  {73}},\ \bibinfo {pages} {053817} (\bibinfo {year} {2006})}\BibitemShut
  {NoStop}%
\bibitem [{\citenamefont {Dalla~Torre}\ \emph {et~al.}(2013)\citenamefont
  {Dalla~Torre}, \citenamefont {Otterbach}, \citenamefont {Demler},
  \citenamefont {Vuletic},\ and\ \citenamefont
  {Lukin}}]{PhysRevLett.110.120402}%
  \BibitemOpen
  \bibfield  {author} {\bibinfo {author} {\bibfnamefont {E.~G.}\ \bibnamefont
  {Dalla~Torre}}, \bibinfo {author} {\bibfnamefont {J.}~\bibnamefont
  {Otterbach}}, \bibinfo {author} {\bibfnamefont {E.}~\bibnamefont {Demler}},
  \bibinfo {author} {\bibfnamefont {V.}~\bibnamefont {Vuletic}},\ and\ \bibinfo
  {author} {\bibfnamefont {M.~D.}\ \bibnamefont {Lukin}},\ }\bibfield  {title}
  {\bibinfo {title} {Dissipative preparation of spin squeezed atomic ensembles
  in a steady state},\ }\href {https://doi.org/10.1103/PhysRevLett.110.120402}
  {\bibfield  {journal} {\bibinfo  {journal} {Phys. Rev. Lett.}\ }\textbf
  {\bibinfo {volume} {110}},\ \bibinfo {pages} {120402} (\bibinfo {year}
  {2013})}\BibitemShut {NoStop}%
\bibitem [{\citenamefont {Bennett}\ \emph {et~al.}(2013)\citenamefont
  {Bennett}, \citenamefont {Yao}, \citenamefont {Otterbach}, \citenamefont
  {Zoller}, \citenamefont {Rabl},\ and\ \citenamefont
  {Lukin}}]{PhysRevLett.110.156402}%
  \BibitemOpen
  \bibfield  {author} {\bibinfo {author} {\bibfnamefont {S.~D.}\ \bibnamefont
  {Bennett}}, \bibinfo {author} {\bibfnamefont {N.~Y.}\ \bibnamefont {Yao}},
  \bibinfo {author} {\bibfnamefont {J.}~\bibnamefont {Otterbach}}, \bibinfo
  {author} {\bibfnamefont {P.}~\bibnamefont {Zoller}}, \bibinfo {author}
  {\bibfnamefont {P.}~\bibnamefont {Rabl}},\ and\ \bibinfo {author}
  {\bibfnamefont {M.~D.}\ \bibnamefont {Lukin}},\ }\bibfield  {title} {\bibinfo
  {title} {Phonon-induced spin-spin interactions in diamond nanostructures:
  Application to spin squeezing},\ }\href
  {https://doi.org/10.1103/PhysRevLett.110.156402} {\bibfield  {journal}
  {\bibinfo  {journal} {Phys. Rev. Lett.}\ }\textbf {\bibinfo {volume} {110}},\
  \bibinfo {pages} {156402} (\bibinfo {year} {2013})}\BibitemShut {NoStop}%
\bibitem [{\citenamefont {{S{\o}rensen}}\ \emph {et~al.}(2001)\citenamefont
  {{S{\o}rensen}}, \citenamefont {{Duan}}, \citenamefont {{Cirac}},\ and\
  \citenamefont {{Zoller}}}]{Natur-409-63}%
  \BibitemOpen
  \bibfield  {author} {\bibinfo {author} {\bibfnamefont {A.}~\bibnamefont
  {{S{\o}rensen}}}, \bibinfo {author} {\bibfnamefont {L.~M.}\ \bibnamefont
  {{Duan}}}, \bibinfo {author} {\bibfnamefont {J.~I.}\ \bibnamefont
  {{Cirac}}},\ and\ \bibinfo {author} {\bibfnamefont {P.}~\bibnamefont
  {{Zoller}}},\ }\bibfield  {title} {\bibinfo {title} {{Many-particle
  entanglement with Bose-Einstein condensates}},\ }\href
  {https://doi.org/10.1038/35051038} {\bibfield  {journal} {\bibinfo  {journal}
  {\nat}\ }\textbf {\bibinfo {volume} {409}},\ \bibinfo {pages} {63} (\bibinfo
  {year} {2001})}\BibitemShut {NoStop}%
\bibitem [{\citenamefont {Burek}\ \emph {et~al.}(2012)\citenamefont {Burek},
  \citenamefont {de~Leon}, \citenamefont {Shields}, \citenamefont {Hausmann},
  \citenamefont {Chu}, \citenamefont {Quan}, \citenamefont {Zibrov},
  \citenamefont {Park}, \citenamefont {Lukin},\ and\ \citenamefont
  {Lon\v{c}ar}}]{domaind-Burek}%
  \BibitemOpen
  \bibfield  {author} {\bibinfo {author} {\bibfnamefont {M.~J.}\ \bibnamefont
  {Burek}}, \bibinfo {author} {\bibfnamefont {N.~P.}\ \bibnamefont {de~Leon}},
  \bibinfo {author} {\bibfnamefont {B.~J.}\ \bibnamefont {Shields}}, \bibinfo
  {author} {\bibfnamefont {B.~J.~M.}\ \bibnamefont {Hausmann}}, \bibinfo
  {author} {\bibfnamefont {Y.}~\bibnamefont {Chu}}, \bibinfo {author}
  {\bibfnamefont {Q.}~\bibnamefont {Quan}}, \bibinfo {author} {\bibfnamefont
  {A.~S.}\ \bibnamefont {Zibrov}}, \bibinfo {author} {\bibfnamefont
  {H.}~\bibnamefont {Park}}, \bibinfo {author} {\bibfnamefont {M.~D.}\
  \bibnamefont {Lukin}},\ and\ \bibinfo {author} {\bibfnamefont
  {M.}~\bibnamefont {Lon\v{c}ar}},\ }\bibfield  {title} {\bibinfo {title}
  {Free-standing mechanical and photonic nanostructures in single-crystal
  diamond},\ }\href {https://doi.org/10.1021/nl302541e} {\bibfield  {journal}
  {\bibinfo  {journal} {Nano Lett.}\ }\textbf {\bibinfo {volume} {12}},\
  \bibinfo {pages} {6084} (\bibinfo {year} {2012})}\BibitemShut {NoStop}%
\bibitem [{\citenamefont {{Tao}}\ \emph {et~al.}(2014)\citenamefont {{Tao}},
  \citenamefont {{Boss}}, \citenamefont {{Moores}},\ and\ \citenamefont
  {{Degen}}}]{2014NC-5-3638T}%
  \BibitemOpen
  \bibfield  {author} {\bibinfo {author} {\bibfnamefont {Y.}~\bibnamefont
  {{Tao}}}, \bibinfo {author} {\bibfnamefont {J.~M.}\ \bibnamefont {{Boss}}},
  \bibinfo {author} {\bibfnamefont {B.~A.}\ \bibnamefont {{Moores}}},\ and\
  \bibinfo {author} {\bibfnamefont {C.~L.}\ \bibnamefont {{Degen}}},\
  }\bibfield  {title} {\bibinfo {title} {{Single-crystal diamond nanomechanical
  resonators with quality factors exceeding one million}},\ }\href
  {https://doi.org/10.1038/ncomms4638} {\bibfield  {journal} {\bibinfo
  {journal} {Nat. Commun.}\ }\textbf {\bibinfo {volume} {5}},\ \bibinfo {eid}
  {3638} (\bibinfo {year} {2014})}\BibitemShut {NoStop}%
\bibitem [{\citenamefont {{Ovartchaiyapong}}\ \emph {et~al.}(2012)\citenamefont
  {{Ovartchaiyapong}}, \citenamefont {{Pascal}}, \citenamefont {{Myers}},
  \citenamefont {{Lauria}},\ and\ \citenamefont {{Bleszynski
  Jayich}}}]{2012ApPhL.101p3505O}%
  \BibitemOpen
  \bibfield  {author} {\bibinfo {author} {\bibfnamefont {P.}~\bibnamefont
  {{Ovartchaiyapong}}}, \bibinfo {author} {\bibfnamefont {L.~M.~A.}\
  \bibnamefont {{Pascal}}}, \bibinfo {author} {\bibfnamefont {B.~A.}\
  \bibnamefont {{Myers}}}, \bibinfo {author} {\bibfnamefont {P.}~\bibnamefont
  {{Lauria}}},\ and\ \bibinfo {author} {\bibfnamefont {A.~C.}\ \bibnamefont
  {{Bleszynski Jayich}}},\ }\bibfield  {title} {\bibinfo {title} {{High quality
  factor single-crystal diamond mechanical resonators}},\ }\href
  {https://doi.org/10.1063/1.4760274} {\bibfield  {journal} {\bibinfo
  {journal} {Appl. Phys. Lett.}\ }\textbf {\bibinfo {volume} {101}},\ \bibinfo
  {eid} {163505} (\bibinfo {year} {2012})}\BibitemShut {NoStop}%
\bibitem [{\citenamefont {{Est{\`e}ve}}\ \emph {et~al.}(2008)\citenamefont
  {{Est{\`e}ve}}, \citenamefont {{Gross}}, \citenamefont {{Weller}},
  \citenamefont {{Giovanazzi}},\ and\ \citenamefont
  {{Oberthaler}}}]{2008Natur-409-63S}%
  \BibitemOpen
  \bibfield  {author} {\bibinfo {author} {\bibfnamefont {J.}~\bibnamefont
  {{Est{\`e}ve}}}, \bibinfo {author} {\bibfnamefont {C.}~\bibnamefont
  {{Gross}}}, \bibinfo {author} {\bibfnamefont {A.}~\bibnamefont {{Weller}}},
  \bibinfo {author} {\bibfnamefont {S.}~\bibnamefont {{Giovanazzi}}},\ and\
  \bibinfo {author} {\bibfnamefont {M.~K.}\ \bibnamefont {{Oberthaler}}},\
  }\bibfield  {title} {\bibinfo {title} {{Squeezing and entanglement in a
  Bose-Einstein condensate}},\ }\href {https://doi.org/10.1038/nature07332}
  {\bibfield  {journal} {\bibinfo  {journal} {Nature (London)}\ }\textbf
  {\bibinfo {volume} {455}},\ \bibinfo {pages} {1216} (\bibinfo {year}
  {2008})}\BibitemShut {NoStop}%
\bibitem [{\citenamefont {{Gross}}\ \emph {et~al.}(2010)\citenamefont
  {{Gross}}, \citenamefont {{Zibold}}, \citenamefont {{Nicklas}}, \citenamefont
  {{Est{\`e}ve}},\ and\ \citenamefont {{Oberthaler}}}]{2010Natur.464.1165G}%
  \BibitemOpen
  \bibfield  {author} {\bibinfo {author} {\bibfnamefont {C.}~\bibnamefont
  {{Gross}}}, \bibinfo {author} {\bibfnamefont {T.}~\bibnamefont {{Zibold}}},
  \bibinfo {author} {\bibfnamefont {E.}~\bibnamefont {{Nicklas}}}, \bibinfo
  {author} {\bibfnamefont {J.}~\bibnamefont {{Est{\`e}ve}}},\ and\ \bibinfo
  {author} {\bibfnamefont {M.~K.}\ \bibnamefont {{Oberthaler}}},\ }\bibfield
  {title} {\bibinfo {title} {{Nonlinear atom interferometer surpasses classical
  precision limit}},\ }\href {https://doi.org/10.1038/nature08919} {\bibfield
  {journal} {\bibinfo  {journal} {\nat}\ }\textbf {\bibinfo {volume} {464}},\
  \bibinfo {pages} {1165} (\bibinfo {year} {2010})}\BibitemShut {NoStop}%
\bibitem [{\citenamefont {{Riedel}}\ \emph {et~al.}(2010)\citenamefont
  {{Riedel}}, \citenamefont {{B{\"o}hi}}, \citenamefont {{Li}}, \citenamefont
  {{H{\"a}nsch}}, \citenamefont {{Sinatra}},\ and\ \citenamefont
  {{Treutlein}}}]{2010Natur.464.1170R}%
  \BibitemOpen
  \bibfield  {author} {\bibinfo {author} {\bibfnamefont {M.~F.}\ \bibnamefont
  {{Riedel}}}, \bibinfo {author} {\bibfnamefont {P.}~\bibnamefont
  {{B{\"o}hi}}}, \bibinfo {author} {\bibfnamefont {Y.}~\bibnamefont {{Li}}},
  \bibinfo {author} {\bibfnamefont {T.~W.}\ \bibnamefont {{H{\"a}nsch}}},
  \bibinfo {author} {\bibfnamefont {A.}~\bibnamefont {{Sinatra}}},\ and\
  \bibinfo {author} {\bibfnamefont {P.}~\bibnamefont {{Treutlein}}},\
  }\bibfield  {title} {\bibinfo {title} {{Atom-chip-based generation of
  entanglement for quantum metrology}},\ }\href
  {https://doi.org/10.1038/nature08988} {\bibfield  {journal} {\bibinfo
  {journal} {\nat}\ }\textbf {\bibinfo {volume} {464}},\ \bibinfo {pages}
  {1170} (\bibinfo {year} {2010})}\BibitemShut {NoStop}%
\bibitem [{\citenamefont {Takeuchi}\ \emph {et~al.}(2005)\citenamefont
  {Takeuchi}, \citenamefont {Ichihara}, \citenamefont {Takano}, \citenamefont
  {Kumakura}, \citenamefont {Yabuzaki},\ and\ \citenamefont
  {Takahashi}}]{PhysRevLett.94.023003}%
  \BibitemOpen
  \bibfield  {author} {\bibinfo {author} {\bibfnamefont {M.}~\bibnamefont
  {Takeuchi}}, \bibinfo {author} {\bibfnamefont {S.}~\bibnamefont {Ichihara}},
  \bibinfo {author} {\bibfnamefont {T.}~\bibnamefont {Takano}}, \bibinfo
  {author} {\bibfnamefont {M.}~\bibnamefont {Kumakura}}, \bibinfo {author}
  {\bibfnamefont {T.}~\bibnamefont {Yabuzaki}},\ and\ \bibinfo {author}
  {\bibfnamefont {Y.}~\bibnamefont {Takahashi}},\ }\bibfield  {title} {\bibinfo
  {title} {Spin squeezing via one-axis twisting with coherent light},\ }\href
  {https://doi.org/10.1103/PhysRevLett.94.023003} {\bibfield  {journal}
  {\bibinfo  {journal} {Phys. Rev. Lett.}\ }\textbf {\bibinfo {volume} {94}},\
  \bibinfo {pages} {023003} (\bibinfo {year} {2005})}\BibitemShut {NoStop}%
\bibitem [{\citenamefont {Leroux}\ \emph {et~al.}(2010)\citenamefont {Leroux},
  \citenamefont {Schleier-Smith},\ and\ \citenamefont
  {Vuleti\ifmmode~\acute{c}\else \'{c}\fi{}}}]{PhysRevLett.104.073602}%
  \BibitemOpen
  \bibfield  {author} {\bibinfo {author} {\bibfnamefont {I.~D.}\ \bibnamefont
  {Leroux}}, \bibinfo {author} {\bibfnamefont {M.~H.}\ \bibnamefont
  {Schleier-Smith}},\ and\ \bibinfo {author} {\bibfnamefont {V.}~\bibnamefont
  {Vuleti\ifmmode~\acute{c}\else \'{c}\fi{}}},\ }\bibfield  {title} {\bibinfo
  {title} {Implementation of cavity squeezing of a collective atomic spin},\
  }\href {https://doi.org/10.1103/PhysRevLett.104.073602} {\bibfield  {journal}
  {\bibinfo  {journal} {Phys. Rev. Lett.}\ }\textbf {\bibinfo {volume} {104}},\
  \bibinfo {pages} {073602} (\bibinfo {year} {2010})}\BibitemShut {NoStop}%
\bibitem [{\citenamefont {Schleier-Smith}\ \emph {et~al.}(2010)\citenamefont
  {Schleier-Smith}, \citenamefont {Leroux},\ and\ \citenamefont
  {Vuleti\ifmmode~\acute{c}\else \'{c}\fi{}}}]{PhysRevA.81.021804}%
  \BibitemOpen
  \bibfield  {author} {\bibinfo {author} {\bibfnamefont {M.~H.}\ \bibnamefont
  {Schleier-Smith}}, \bibinfo {author} {\bibfnamefont {I.~D.}\ \bibnamefont
  {Leroux}},\ and\ \bibinfo {author} {\bibfnamefont {V.}~\bibnamefont
  {Vuleti\ifmmode~\acute{c}\else \'{c}\fi{}}},\ }\bibfield  {title} {\bibinfo
  {title} {Squeezing the collective spin of a dilute atomic ensemble by cavity
  feedback},\ }\href {https://doi.org/10.1103/PhysRevA.81.021804} {\bibfield
  {journal} {\bibinfo  {journal} {Phys. Rev. A}\ }\textbf {\bibinfo {volume}
  {81}},\ \bibinfo {pages} {021804} (\bibinfo {year} {2010})}\BibitemShut
  {NoStop}%
\bibitem [{\citenamefont {Julsgaard}\ \emph {et~al.}(2013)\citenamefont
  {Julsgaard}, \citenamefont {Grezes}, \citenamefont {Bertet},\ and\
  \citenamefont {M\o{}lmer}}]{PhysRevLett.110.250503}%
  \BibitemOpen
  \bibfield  {author} {\bibinfo {author} {\bibfnamefont {B.}~\bibnamefont
  {Julsgaard}}, \bibinfo {author} {\bibfnamefont {C.}~\bibnamefont {Grezes}},
  \bibinfo {author} {\bibfnamefont {P.}~\bibnamefont {Bertet}},\ and\ \bibinfo
  {author} {\bibfnamefont {K.}~\bibnamefont {M\o{}lmer}},\ }\bibfield  {title}
  {\bibinfo {title} {Quantum memory for microwave photons in an inhomogeneously
  broadened spin ensemble},\ }\href
  {https://doi.org/10.1103/PhysRevLett.110.250503} {\bibfield  {journal}
  {\bibinfo  {journal} {Phys. Rev. Lett.}\ }\textbf {\bibinfo {volume} {110}},\
  \bibinfo {pages} {250503} (\bibinfo {year} {2013})}\BibitemShut {NoStop}%
\bibitem [{\citenamefont {Schuetz}\ \emph {et~al.}(2015)\citenamefont
  {Schuetz}, \citenamefont {Kessler}, \citenamefont {Giedke}, \citenamefont
  {Vandersypen}, \citenamefont {Lukin},\ and\ \citenamefont
  {Cirac}}]{PhysRevX.5.031031}%
  \BibitemOpen
  \bibfield  {author} {\bibinfo {author} {\bibfnamefont {M.~J.~A.}\
  \bibnamefont {Schuetz}}, \bibinfo {author} {\bibfnamefont {E.~M.}\
  \bibnamefont {Kessler}}, \bibinfo {author} {\bibfnamefont {G.}~\bibnamefont
  {Giedke}}, \bibinfo {author} {\bibfnamefont {L.~M.~K.}\ \bibnamefont
  {Vandersypen}}, \bibinfo {author} {\bibfnamefont {M.~D.}\ \bibnamefont
  {Lukin}},\ and\ \bibinfo {author} {\bibfnamefont {J.~I.}\ \bibnamefont
  {Cirac}},\ }\bibfield  {title} {\bibinfo {title} {Universal quantum
  transducers based on surface acoustic waves},\ }\href
  {https://doi.org/10.1103/PhysRevX.5.031031} {\bibfield  {journal} {\bibinfo
  {journal} {Phys. Rev. X}\ }\textbf {\bibinfo {volume} {5}},\ \bibinfo {pages}
  {031031} (\bibinfo {year} {2015})}\BibitemShut {NoStop}%
\bibitem [{\citenamefont {Johansson}\ \emph {et~al.}(2013)\citenamefont
  {Johansson}, \citenamefont {Nation},\ and\ \citenamefont {Nori}}]{CPC}%
  \BibitemOpen
  \bibfield  {author} {\bibinfo {author} {\bibfnamefont {J.~R.}\ \bibnamefont
  {Johansson}}, \bibinfo {author} {\bibfnamefont {P.~D.}\ \bibnamefont
  {Nation}},\ and\ \bibinfo {author} {\bibfnamefont {F.}~\bibnamefont {Nori}},\
  }\bibfield  {title} {\bibinfo {title} {Qutip 2: A python framework for the
  dynamics of open quantum systems},\ }\href
  {https://doi.org/10.1016/j.cpc.2012.11.019} {\bibfield  {journal} {\bibinfo
  {journal} {Comput. Phys. Commun.}\ }\textbf {\bibinfo {volume} {184}},\
  \bibinfo {pages} {1234} (\bibinfo {year} {2013})}\BibitemShut {NoStop}%
\end{thebibliography}

%

\end{document}